A NUMERICAL SOLUTION OF THE TIME-DEPENDENT NEUTRON TRANSPORT EQUATION USING A CHARACTERISTIC METHOD. APPLICATIONS TO ICF AND TO HYBRID FISSION-FUSION SYSTEMS

Dalton Ellery Girão Barroso
*Military Institute of Engineering*
*Department of Nuclear Engineering*
*Emails: daltongirao@yahoo.com.br, Dalton@ime.eb.br*

**ABSTRACT**

*In this work we present a solution of the one-dimensional spherical symmetric time-dependent neutron transport equation (written for a moving system in lagrangian coordinates) by using the characteristic method. One of the objectives is to overcome the negative flux problem that arises when the system is very opaque and the angular neutron flux can become negative when it is extrapolated in spatial meshes— as, for example, in diamond scheme adopted in many codes. Although there are recipes to overcome this problem, such procedures can degrade the numerical solution if repeated many times.*

*In most common time discretization of the time-dependent neutron transport equation, there is a term, $1/v\Delta t$, summed to total (or transport) cross sections that can increase enormously the opacity of the medium. This certainly gets worse when there is a combination of lower* neutron velocity (in a moderated system) *and a very small time-step — a very common situation in ICF (inertial confinement fusion).*

*The solution presented here can be easily coupled to radiation-hydrodynamics equations, but it is necessary an additional term to maintain neutron conservation in a moving system in lagrangian coordinates. Energy multigroup method and a former $S_N$ method to deal with the angular variable are used, with the assumption of isotropic scattering and the transport cross sections approximation. An artifice is employed for emulating the neutron upscattering when neutron energy is much lower than the temperature (in kT units) of the medium. The consistency of the numerical solution is checked by making at each time-step the balance of neutrons in the system.*

*Two examples of applications are shown using a neutronic-radiation-hydrodynamic code to which the solution here presented was incorporated: one consists of a heterogeneous pellet of DT (deuterium-tritium) tamped by an highly-enriched uranium or plutonium (a symbiotic fusion-fission system); and the other is a very complex and also a symbiotic fission-fusion-fission system composed by layers of the thermonuclear fuels LiDT,*

*LiD and a highly-supercritical fission fuel. The last is considered an extreme case for testing the time-dependent neutron transport solution presented here.*

## 1. Introduction

In many ICF (inertial confinement fusion) simulations, neutron transport and the energy deposited by neutrons on the target pellet are disregarded because generally the neutron mean free path is much greater than or of the order of the dimension of the pellet; in these cases neutrons are assumed to escape freely from the pellet. However, in other cases this is not true and the neutron transport and the influence of neutron energy deposition on the dynamic behavior of the pellet have to be taken into account. Many use simplified models,[1-4] while others couple the radiation-hydrodynamic equations to a consecrated time-dependent transport code, as the TIMEX code,[5-7] or even use a more rigorous model taking into account all additional terms that appear in neutron transport equation when this transport is performed in a moving system.[8]

It is very difficult to testify the consistency of these solutions and schemes in all cases because we have no numerical details about them and the authors do not provide exhaustive tests employing them (at least as far as I know).

One of the most difficult problems in time-dependent neutron transport solution concerns the generation of negative flux due to a great opacity provided by an additional term summed to total neutron cross section that arises when the transport equation is discretized in time, as we will see forward.

The numerical solution developed here employs the well known characteristic method to overcome this problem. This solution has proved to be practically free of the negative flux problem.

Here we followed the scheme developed by Carlson.[9] (A similar scheme is presented in the Clark-Hansen book[10].)

## 2. The time-dependent neutron transport equation in a moving system in lagrangian coordinates.

Normally, the neutron transport equation is established making the neutron balance (at point r, energy E, angular direction $\hat{\Omega}$ and time t) in an infitesimal volume element:

$$\frac{\partial}{\partial t}\int_{\Delta V}\mathbf{N}dV = \text{[(rate of neutron entering in } \Delta V\text{)+(rate of neutrons produced in } \Delta V\text{)]-}$$

$$\text{-[(rate of neutrons leaked out from } \Delta V\text{)+(rate of neutrons disappearing in } \Delta V\text{)]} , \qquad (1)$$

where $N(r,E,\hat{\Omega},t)$ is the angular neutron density.

When the volume element is fixed, i.e., the medium is static (as in nuclear reactors), the time derivative can be displaced into the integral. But for a moving medium (as the fissile mass in a nuclear explosion, for example) this displacement can not be made. However, in lagrangian coordinates, where the mass points move with the material, the mass element $\Delta m$ inside the volume $\Delta V$ does not vary, only the density varies. Thus, the time derivative can be put into the integral by replacing dV by $dm/\rho$, and then we have:

$$\frac{\partial}{\partial t}\int_{\Delta V}\mathbf{N}dV = \frac{\partial}{\partial t}\int\frac{N}{\rho}dm = \int\frac{\partial(N/\rho)}{\partial t}dm = \int\frac{\partial(N/\rho)}{\partial t}\rho dV = \int[\frac{\partial N}{\partial t} - N\frac{\partial \ln\rho}{\partial t}]dV \ . \qquad (2)$$

Note that this procedure creates an additional term in the transport equation, given by $-N\partial\ln\rho/\partial t$. Numerically, to maintain neutron conservation, this term makes a correction in the angular neutron flux (there is, in the angular neutron density) in the meshes, when these meshes move (contracting or expanding) due to hydrodynamic calculations. Thus, with this additional term the transport equation is given by:

$$\frac{1}{v}\frac{\partial\Psi(r,E,\hat{\Omega},t)}{\partial t} - \frac{1}{v}\Psi(r,E,\hat{\Omega},t)\frac{\partial\ln\rho}{\partial t} + \hat{\Omega}.\nabla\Psi(r,E,\hat{\Omega},t) + \Sigma_t(r,E,t)\Psi(r,E,\hat{\Omega},t)$$

$$= \int_{4\pi}d\hat{\Omega}'\int_0^\infty dE'\Sigma_s(r,E'\Rightarrow E,\hat{\Omega}'\Rightarrow\hat{\Omega},t)\,\Psi(r,E',\hat{\Omega}',t) \ +$$

$$+\frac{\chi(E)}{4\pi}\int_{4\pi}d\hat{\Omega}'\int_0^\infty dE'\nu(E')\Sigma_f(r,E',t)\,\Psi(r,E',\hat{\Omega}',t) + Q(r,E,\hat{\Omega},t) \ . \qquad (3)$$

In one-dimensional spherical symmetry and admitting only isotropic scattering (with correction given by using the transport cross sections), the Eq.(3) is reduced to:

$$\frac{\partial}{v\partial t}\Psi(r,E,\mu,t) + \mu\frac{\partial}{\partial r}\Psi(r,E,\mu,t) + \frac{1}{r}(1-\mu^2)\frac{\partial}{\partial\mu}\Psi(r,E,\mu,t) +$$

$$+[\Sigma_{tr}(r,E,t) - \frac{1}{v}\frac{\partial\ln\rho}{\partial t}]\Psi(r,E,\mu,t) = S(r,E,\mu,t), \qquad (4)$$

where the source term is given by:

$$S(r,E,\mu,t) = \frac{1}{2}\int_0^\infty\Sigma_s(r,E'\Rightarrow E,t)\int_{-1}^1\Psi(r,E',\mu,t)d\mu'dE' +$$

$$+\frac{\chi(E)}{2}\int_0^\infty\nu(E')\Sigma_f(r,E',t)\int_{-1}^1\Psi(r,E',\mu',t)d\mu'dE' + Q(r,E,\mu,t). \qquad (5)$$

The terms have their usual interpretation ($\mu$ is the cosine of angle between radius r and neutron direction). Note that the additional hydrodynamic term was grouped with transport cross section. When the system is fixed, $\partial\rho/\partial t = 0$ and we have the usual transport equation.

The external neutron source is considered to be a fusion source:

$$Q(r,E,\hat{\Omega},t) = \frac{1}{4\pi}[\chi_{DD}(E) < v\sigma_{DD} > n_D^2(r,t)/4 + \chi_{DT}(E) < v\sigma_{DT} > n_D(r,t)n_T(r,t)], \quad (6)$$

where $\chi_{DD}(E)$ and $\chi_{DT}(E)$ are the fusion neutron spectra of D-D and D-T reactions, respectively [$\chi_{DT}$(14.1 MeV)=1 and $\chi_{DD}$(2.45 MeV)=1 and zero for other energies], $< v\sigma >$, the particle velocity times the cross sections of fusion reactions weighted in Maxwellian velocity distribution (there are good temperature-dependent semi-empirical formulas for them[11]), and $n_D$ and $n_T$ are the deuterium and tritium numeric densities. The division by $4\pi$ is due to the fact that fusion neutrons are emitted isotropically (in D-T reaction and in one branch of D-D reactions).

It is important to highlight that other corrections to be made in neutron transport in a moving system[12,13,8] (for example, neutron kinetic energy increases if it is emitted in direction of the motion and vice-versa) are considered to have a minor influence in comparison to the correction term presented here. This means that in our cases here we consider that the neutron velocities are much greater than hydrodynamic velocities of the medium at all time of interest.

**3. A commonly used time discretization of the transport equation**

Consider the Eq.(3) written in the following compact form (disregarding the additional term):

$$\frac{1}{v}\frac{\partial\Psi}{\partial t} + \hat{\Omega}.\nabla\Psi + \Sigma_t\Psi = S, \quad (7)$$

where S represents the right-hand-side source term of Eq.(3).

The time derivative is discretized by the simple formula:

$$\frac{1}{v}\frac{\partial\Psi}{\partial t} = \frac{1}{v}\frac{(\Psi^{n+1} - \Psi^{n})}{\Delta t}.$$

Replacing into Eq.(7) and admitting all the other terms in the advanced time n+1, we have:

$$\frac{\Psi^{n+1} - \Psi^{n}}{v\Delta t} + \hat{\Omega}.\nabla\Psi^{n+1} + \Sigma_t\Psi^{n+1} = S^{n+1}.$$

Rearranging:

$$\hat{\Omega}.\nabla\Psi^{n+1} + (\Sigma_t + \frac{1}{v\Delta t})\Psi^{n+1} = S^{n+1} + \frac{\Psi^n}{v\Delta t}. \tag{8}$$

This equation is exactly equal to the static transport equation since we define the new terms:

$$\Sigma_t^* = \Sigma_t + \frac{1}{v\Delta t}, \tag{9}$$

$$S^{*n+1} = S^{n+1} + \frac{\Psi^n}{v\Delta t}. \tag{10}$$

The expression (9) defines a new total (or transport) cross section with the addition of the term $1/v\Delta t$. In (10) the neutron source (due to scattering, fission and fusion) is added by a term that depends on the angular neutron flux calculated in the previous time-step (and hence already known).

This implicit scheme in time can be solved by iterative method. TIMEX,[7] for example (one of the most known transport codes) uses similar discretization to solve numerically the time-dependent neutron transport equation.

**4. The negative flux problem**

As we mentioned before, the problem is that when the time absorption term $1/v\Delta t$ (we call it so) is much greater than $\Sigma_t$ the system becomes very opaque and, in some numerical scheme, this can generate very negative angular neutron flux in spatial meshes. For example, in diamond scheme, to close the equations, the angular neutron flux is linearly extrapolated in the meshes by the expression (in outward direction - see Fig.1):

$$\Psi_{i+1/2}=2\Psi_i-\Psi_{i-1/2}.$$

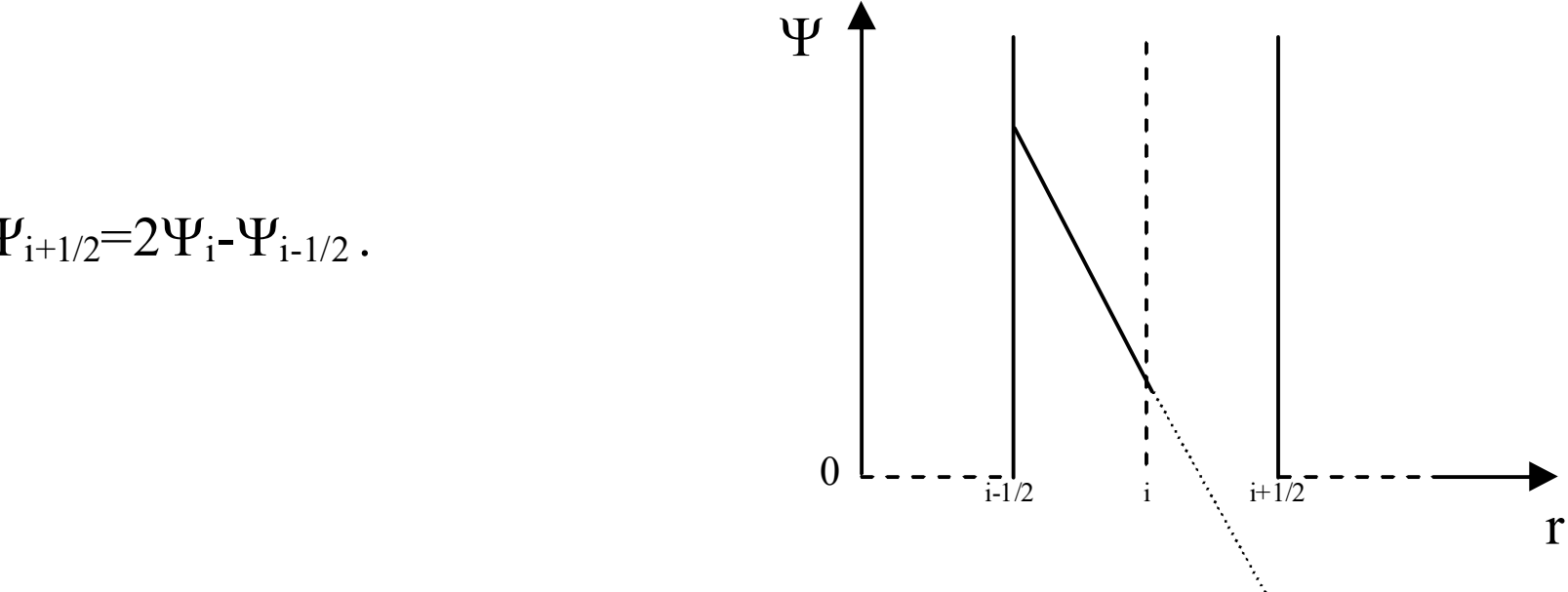

**Figure 1**: Linear extrapolation of neutron flux can generate negative values if the system is very opaque.

$\Psi_{i+1/2}$ will be negative when $\Psi_{i-1/2}>2\Psi_i$. Although there are recipes to circumvent this problem (*negative flux fixup*)[14,15] — for example, zeroing these fluxes and maintaining the neutrons conservation, or using the step or a weight numeric schemes —, the process used, if repeated many times, can lead to degradation of the stability and precision of the numerical solution.

**5. The characteristic method**

The characteristic method has been known as one of the best numerical methods to overcome the negative flux problem.

In what follows, we will present the numerical solution of the time-dependent neutron transport equation (4) by a characteristic method using the Carlson scheme.[9] (In the angular variable we will also use an old $S_N$ method developed by Carlson,[16] with quadrature trapezoidal.)

In the original $S_N$ method, where trapezoidal quadrature is used, the angular variable $\mu$ ($\mu$ varies from -1 to 1) is discretized into J number of interval directions (N=J+1 directions) given by: $\mu_j=-1+2j/J$ (j=0,1,...J) ($\Delta_j=2/J$). The neutron angular flux is given by:

$$\Psi(r,E,\mu,t)=\frac{\mu-\mu_{j-1}}{\mu_j-\mu_{j-1}}\Psi(r,E,\mu_j,t)+\frac{\mu_j-\mu}{\mu_j-\mu_{j-1}}\Psi(r,E,\mu_{j-1},t) \qquad (11)$$

in the interval $\mu_{j-1}\le\mu\le\mu_j$. Considering that the external source does not depend on the angular variable (for example, fusion neutrons are emitted isotropically), S in Eq.(5) does not depend on $\mu$ (this dependence could be easily taken into account). Replacing (11) into (4) and integrating in $\mu$ from $\mu_{j-1}$ to $\mu_j$, we have (in multigroup notation):

$$\left[\frac{1}{v_g}\frac{\partial}{\partial t}+a_j\frac{\partial}{\partial r}+\frac{b_j}{r}+\Sigma^*_{trg}(r,t)\right]\Psi_{g,j}(r,t)+\left[\frac{1}{v_g}\frac{\partial}{\partial t}+\bar{a}_j\frac{\partial}{\partial r}-\frac{b_j}{r}+\Sigma^*_{trg}(r,t)\right]\Psi_{g,j-1}(r,t)=$$
$$=c_k S_g(r,t), \qquad (12)$$

where: $\Psi_{g,j}(r,t)=\Psi_g(r,\mu_j,t)$;

$$\Sigma^*_{trg}(r,t)=\Sigma_{trg}(r,t)-\frac{1}{v_g}\frac{\partial\ln\rho}{\partial t};$$

$a_j=(2\mu_j+\mu_{j-1})/3$;

$\bar{a}_j=(\mu_j+2\mu_{j-1})/3$;

$$b_j=\frac{2}{3}(3-\mu_j^2-\mu_j\mu_{j-1}-\mu_{j-1}^2)/\Delta_j;$$

$j\neq0$, $c_k=2$; $j=0$, $c_k=1$, $a_0=-1$, $b_0=0$ and $\Psi_{g,-1}(r,t)=0$.

The solution of Eq.(12) in space and time, using the characteristic method, is performed by making the solution along the angular flux direction specified in the space-time diagram of Figure 2. The solution of directional neutron flux at the point C depends on its known values at the points A, B and D. Two cases are considered depending on the following conditions (see the notations of the variables forward):

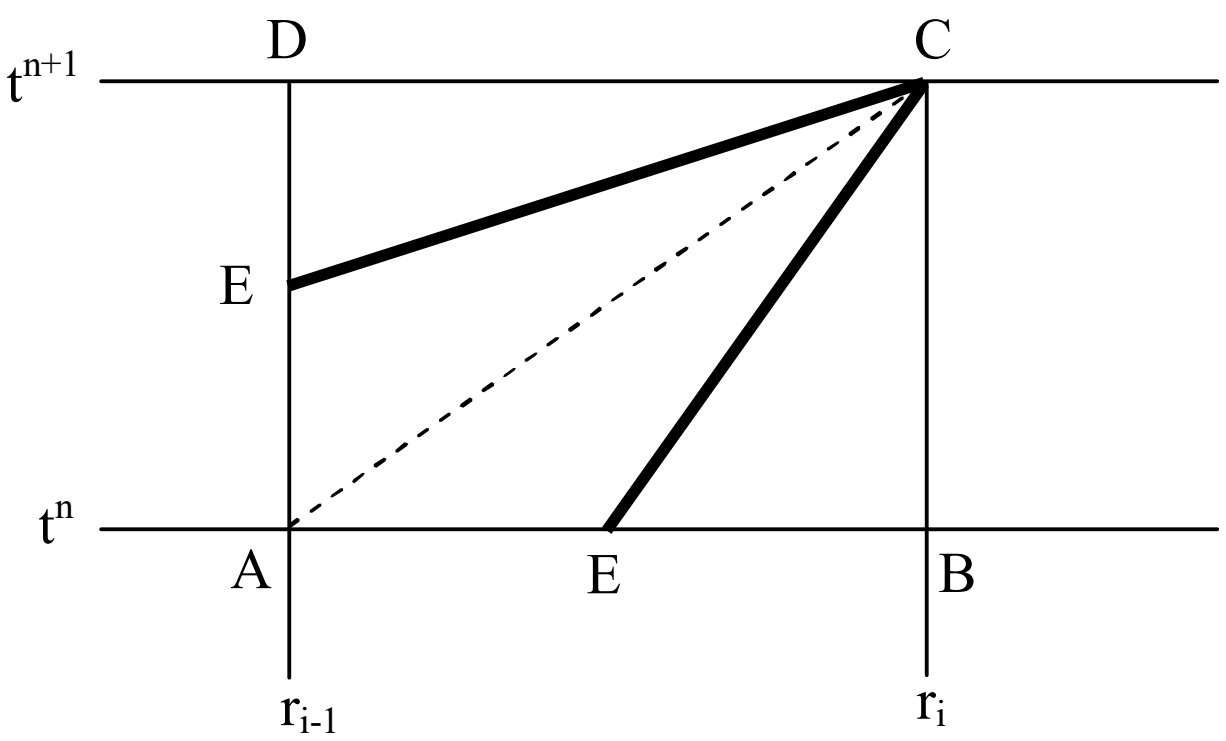

**Figure 2:** Directional flux of neutrons in the space-time diagram for μ>0.

a) $\left|a_j\right| < \dfrac{\Delta_i}{v_g \Delta^{n+1/2}}$ ($\Delta_i = r_i - r_{i-1}$)

In this case, the neutron path EC crosses the line AB in the time-step $\Delta^{n+1/2}=t^{n+1}-t^n$; the time derivative of the angular neutron flux is made along the line BC, while the space derivative, along the line AB (this means that the total derivative is performed along the characteristic line EC that crosses AB, assuming linear variation between $r_i$ and $r_{i-1}$):

$$\frac{\partial \Psi_{g,j}(r,t)}{\partial t} = \frac{\Psi_{g,j}(r_i, t^{n+1}) - \Psi_{g,j}(r_i, t^n)}{t^{n+1} - t^n},$$

$$\frac{\partial \Psi_{g,j}(r,t)}{\partial r} = \frac{\Psi_{g,j}(r_i, t^n) - \Psi_{g,j}(r_{i-1}, t^n)}{r_i - r_{i-1}}.$$

b) $\left|a_j\right| > \dfrac{\Delta_i}{v_g \Delta^{n+1/2}}$

In this second case, the neutron path EC crosses the line AD in the time-step $\Delta^{n+/2}=t^{n+1}-t^n$; the time derivative is performed along AD line and the spatial, along DC (the total derivative being now along EC that crosses the line AD):

$$\frac{\partial \Psi_{g,j}(r,t)}{\partial t} = \frac{\Psi_{g,j}(r_{i-1}, t^{n+1}) - \Psi_{g,j}(r_{i-1}, t^n)}{t^{n+1} - t^n},$$

$$\frac{\partial \Psi_{g,j}(r,t)}{\partial r} = \frac{\Psi_{g,j}(r_i, t^{n+1}) - \Psi_{g,j}(r_{i-1}, t^{n+1})}{r_i - r_{i-1}}.$$

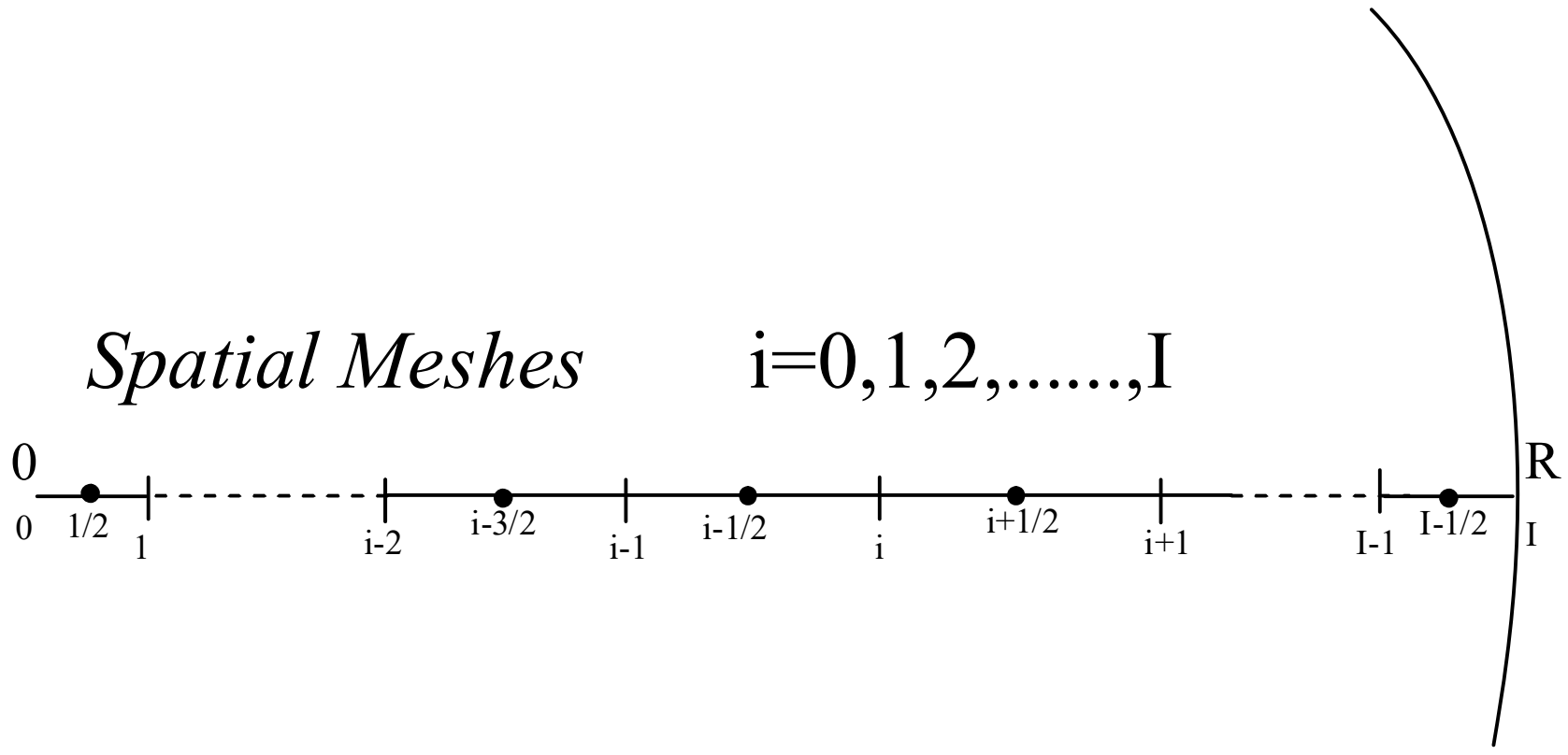

**Figure 3:** Meshes in spatial discretization of transport equation. (The same scheme is used to solve the radiation-hydrodynamic equations in our code. In lagrangian coordinates the meshes move with the material.)

Before presenting the final numerical scheme used in the characteristic method, the following notation is used:

i=1,2...I – numeric index of the spatial variable; the index i-1/2 refers to the centre of the intervals (see Fig.3);

g=1,2...G – numeric index of energy groups;

j=1,2,...J+1 – numeric index of the angular variable (the j was shifted by 1 to avoid the index 0); J is the number of angular intervals;

n=1,2...N – numeric index of the time variable;

$r_i^n$ – radius of the spatial interval i; (in lagrangian coordinates it moves with the mass points;

$\Psi_{g,i,j}^n$ - angular neutron flux in the energy group g, spatial interval i, angular direction j and time n;

$v_g$ - velocity of neutrons in group g;

$\Delta_j = \mu_j - \mu_{j-1}$ - angular interval;

$\Delta_i = r_i^n - r_{i-1}^n$ – spatial interval;

$\Delta^{n+1/2} = t^{n+1} - t^n$ – time-step;

$V_{i-1/2}^n = \frac{4}{3}\pi[(r_i^n)^3 - (r_{i-1}^n)^3]$ - volume of the region between $r_i^n$ e $r_{i-1}^n$.

The boundary conditions at the centre and at the external surface (vacuum

condition) are given by:

$\Psi(0,E,\mu,t)=\Psi(0,E,-\mu,t)$, that is, $\Psi^{n}_{g,1,j}=\Psi^{n}_{g,1,jk}$, $jk=(J+2)-j$, $j>J/2+1$

$\Psi(R,E,\mu,t)=0$ for $\mu\le 0$, or, numerically, $\Psi^{n}_{g,I,j}=0,\ \ j\le J/2+1$

The final numeric solution presented below (in notation used here) was obtained by following the same (and laborious) scheme developed by Carlson[9] (in these ancient times, a horrible notation was used). Similar scheme is presented in the Clark-Hansen book.[10]

**Case a)** $\left|a_j\right|<\dfrac{\Delta_i}{v_g\Delta^{n+1/2}}$, $\mu>0$ ($j>J/2+1$, integration from the centre to the external surface):

$$\Psi^{n+1}_{g,i,j}=\frac{\left|a_j\right|(1-W)[\Psi^{n}_{g,i,j}+\Psi^{n}_{g,i,j-1}-\Psi^{n+1}_{g,i,j-1}]+W.K}{\left|a_j\right|(1-W)+W(\left|a_j\right|+H_i+B_i)},\tag{13}$$

where:

$$H_i=\frac{\Sigma^{*}_{trg,i-1/2}\Delta_i}{2};$$

$$\Sigma^{*}_{trg,i-1/2}=\Sigma^{n+1}_{tr,g,i-1/2}+\frac{\ln(V^{n+1}_{i-1/2}/V^{n}_{i-1/2})}{v_g\Delta^{n+1/2}};^{(*)}$$

$$B_i=\frac{b_j\Delta_i}{2\bar{r}_i};$$

$$\bar{r}_i=(r^{n}_{i}+r^{n}_{i-1})/2;$$

$$W=\frac{\left|a_j\right|v_g\Delta^{n+1/2}}{\Delta_i};$$

$$K=(\left|a_j\right|-H_i-B_i)\Psi^{n}_{g,i-1,j}+(\left|a_j\right|-H_i+B_i)\Psi^{n}_{g,i-1,j-1}-(\left|a_j\right|+H_i-B_i)\Psi^{n+1}_{g,i,j-1}+$$
$$\frac{\beta}{2}(\Psi^{n+1}_{g,i,j-1}+\Psi^{n}_{g,i,j-1}-\Psi^{n+1}_{g,i-1,j-1}-\Psi^{n}_{g,i-1,j-1})+\frac{c_k\Delta_i}{2}(S^{n+1}_{g,i-1/2}+S^{n}_{g,i-1/2}).$$

$$\beta=\left|a_j\right|-\left|\bar{a}_j\right|$$

For $\mu\le 0$ ($j=1,2,...J/2+1$, integration from the surface to the centre), we exchange, in the fluxes $\Psi$s above (including the K term), the index i by i-1 and vice-versa; for $j=1$ ($\mu_1=-1$, direction along the radius r toward the centre), $B_i=0$ and all the $\Psi$s with index j-1 are nulls.

---

(*) If this additional term is treated explicitly, we may simply multiply the angular neutron fluxes by relative volume variation at the beginning of each time-step to maintain neutron conservation.

**Caso b)** $\left|a_j\right| > \dfrac{\Delta_i}{v_g \Delta^{n+1/2}}$, $\mu > 0$:

$$\Psi^{n+1}_{g,i,j} = \frac{\left|a_j\right|(1-1/W)(\Psi^{n+1}_{g,i-1,j} + \Psi^{n+1}_{g,i-1,j-1} - \Psi^{n}_{g,i-1,j} - \Psi^{n}_{g,i-1,j-1}) + K}{\left|a_j\right| + H_i + B_i} . \qquad (14)$$

For $\mu \leq 0$, we exchange, in the same way in the Ψs, the index i by i-1 and vice-versa.

Note that in Eq.(13), by the condition a), W is lower than 1 and the term (1-W) is always positive. In Eq.(14), by the condition b), W is greater than 1 and the term (1-1/W) is always positive as well.

The source term S is given by:

(Zero for h>g; upscattering is not considered)

$$S^{n+1}_{g,i-1/2} = \frac{1}{4\pi}[\sum_{h=1}^{g-1} \Sigma^{n+1}_{h\to g,i-1/2}\phi^{n+1}_{h,i-1/2} + \sum_{h=g}^{G} \Sigma^{n+1}_{h\to g,i-1/2}\phi^{n}_{h,i-1/2}] +$$

$$\frac{1}{4\pi}\chi_g[\sum_{h=1}^{g-1}(\nu\Sigma_f)^{n+1}_{h,i-1/2}\phi^{n+1}_{h,i-1/2} + \sum_{h=g}^{G}(\nu\Sigma_f)^{n+1}_{h,i-1/2}\phi^{n}_{h,i-1/2}] + Q^{n+1}_{g,i-1/2} , \qquad (15)$$

where the neutron flux $\phi$ is calculated by the trapezoidal integration in angle:

$$\phi^{n+1}_{g,i-1/2} = \frac{2\pi}{J}[\Psi^{n+1}_{g,i-1/2,1} + 2\sum_{j=2}^{J}\Psi^{n+1}_{g,i-1/2,j} + \Psi^{n+1}_{g,i-1/2,J+1}];$$

$$\Psi^{n+1}_{g,i-1/2,j} = \frac{\Psi^{n+1}_{g,i,j} + \Psi^{n+1}_{g,i-1,j}}{2} . \qquad (16)$$

For the fusion source we have:

$$Q^{n+1}_{g,i-1/2} = \frac{1}{4\pi}[\chi_{DDg} < v\sigma_{DD} >^{n+1}_{i-1/2} (n^{n+1}_{Di-1/2})^2/4 + \chi_{DTg} < v\sigma_{DT} >^{n+1}_{i-1/2} n^{n+1}_{Di-1/2} n^{n+1}_{Ti-1/2}]. \qquad (17)$$

The neutron current is given in the same way by a trapezoidal integration:

$$J^{n+1}_{g,i} = \frac{2\pi}{J}[\mu_1\Psi^{n+1}_{g,i,1} + 2\sum_{j=2}^{J}\mu_j\Psi^{n+1}_{g,i,j} + \mu_{J+1}\Psi^{n+1}_{g,i,J+1}]. \qquad (18)$$

In sequence of computational loop, the solution above develops at each time $t^n$ (n=1,2,...N), for each energy group $E_g$ (g=1,2,...G), for each angular direction $\mu_j$ (j=1,2,... J+1) and in the space intervals from i=I,I-1,...1 (when $\mu_j \leq 0$) or i=1,2,...I (when $\mu_j > 0$).

The energy deposited by neutrons (via downscattering, mainly in light material) per unit volume in mesh i and time-step n+1 is given, with good approximation, by:

$$\Delta W^{n+1}_{i-1/2} = \sum_{g=2}^{G}\sum_{h=1}^{g-1}[\Sigma^{n+1}_{h\to g,i-1/2}\phi^{n+1}_{h,i-1/2}(\overline{E}_h - \overline{E}_g)]\Delta^{n+1/2}, \quad (19)$$

where $\overline{E}_g = mv_g^2/2$ is the mean energy of group g.

An iterative procedure is used to solve the equations. In the solution for the time-step n+1, the implicit source terms, $S^{n+1}_{g,i-1/2}$, are initially defined as a function of their values in the previous time-step n ( $S^{n+1}_{g,i-1/2} = S^{n}_{g,i-1/2}$ ). This source terms are then recalculated using the news values of the obtained neutron fluxes and so on, the process is repeated, until the values of the fluxes converge to a pre-assigned tolerance. Noted that, as can be seen in Eq.(15), the fluxes calculated for the groups $g'<g$ ($E_{g'}>E_g$) are immediately incorporated into the solution for the group g in the same iteration, thereby accelerating the convergence (two iterations, or even one in many cases, have been enough for the majority of the cases tested by the author).

To assure the consistence and stability of the method and to detect any breakdown of the numerical solution it is necessary to check continuously the conservation of neutrons in the system (this is analog to checking the conservation of energy in hydrodynamic solutions). In neutron balance, the number of neutrons produced (in our case, by fission, fusion and (n,2n) reactions) has to be equal to the number of neutrons present in the system plus the number of neutrons that were absorbed and leaked out, at each time.

## 6. A correction in neutron fluxes due to downscattering to energies lower than the temperature (in keV) of the medium.

Due to extreme temperature variations during the nuclear and thermonuclear explosions (from a few tens of eV to several tens or even hundreds of keV), neutrons that were downscattered to energies much lower than the mean energy of the plasma particles, in a given time during the fission and/or fusion reactions growth, can not stay in these low energy states because collisions with the ions tend to upscatter them up to thermal equilibrium with the ions. Depending on the group structure used in the solution of the multigroup neutron transport equation, such fact has to be taken into account. To overcome in part this problem, without explicitly including the very complex temperature-dependent upscattering process, we apply the following procedure. Whenever there are neutron populations (due to scattering) in energy groups whose upper energy boundaries are lower than the temperature (in keV) of the medium, these neutron populations are automatically transferred to energy group in whose limits the temperature of the medium is located (see Fig.4 for visualization of the process in terms of angular fluxes). Neutrons are conserved in this process.

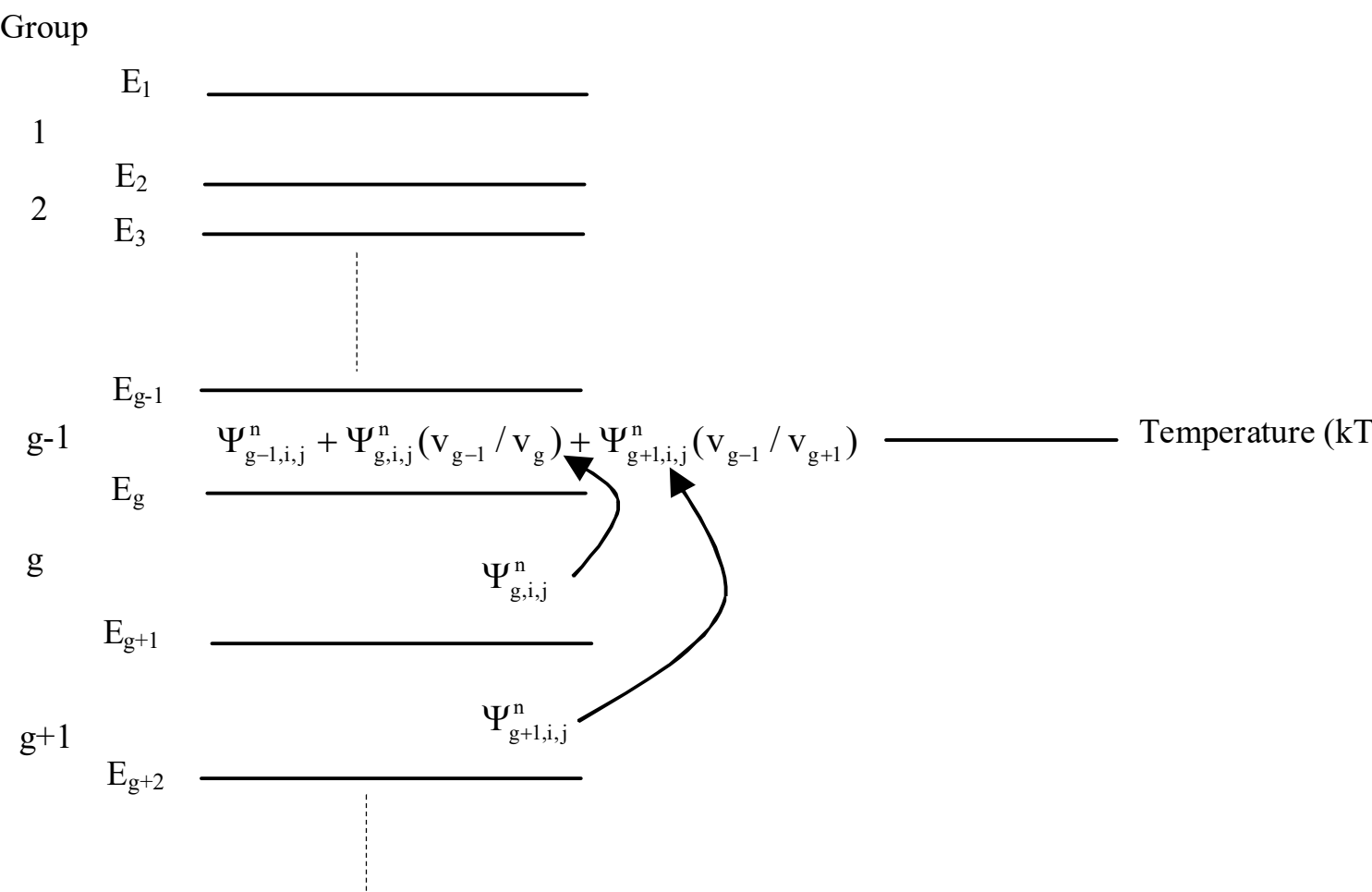

**Figure 4:** Neutrons transfer from groups whose upper energy boundaries are lower than the temperature (in keV) of the medium to the group whose upper energy boundary is immediately greater than kT, emulating, in some way, the upscattering.

## 7. Two cases considered: micro-targets involving fission-fusion processes

In this work we consider two cases involving symbiotic systems composed by fission-fusion targets to test the numerical solution of the time-dependent neutron transport equation presented here. Considerations about the feasibility of the initial conditions of these targets were not performed and have no importance, because our main objective is to test the performance of this solution. Nevertheless, very interesting results were obtained concerning fission-fusion targets performance.

A symbiotic system involving fission-fusion reactions can be realized in two ways: A fusion explosion enhanced by fission reactions in a fissile material incorporated into the thermonuclear target; or a fission explosion of a supercritical fissile material enhanced by fusion reactions in an embodied thermonuclear fuel ignited by this previous fission explosion. In both cases (considered here) the full solution of the time-dependent neutron transport equation is required for simulating the target performance.

### 7.1 A thermonuclear fuel target tamped by a fissile material

In the first case, we considered the target shown in Figure 5. The heterogeneous DT fuels and the initial conditions are the same as those considered in ref.[4]. The added tamper is composed by a layer of 93% enriched uranium or plutonium with varying dimensions. The admitted composition of (reactor grade) plutonium was: $Pu^{238}$(1.3%), $Pu^{239}$(60.3%), $Pu^{240}$(24.3%), $Pu^{241}$(9.1%) and $Pu^{242}$(5.0%).

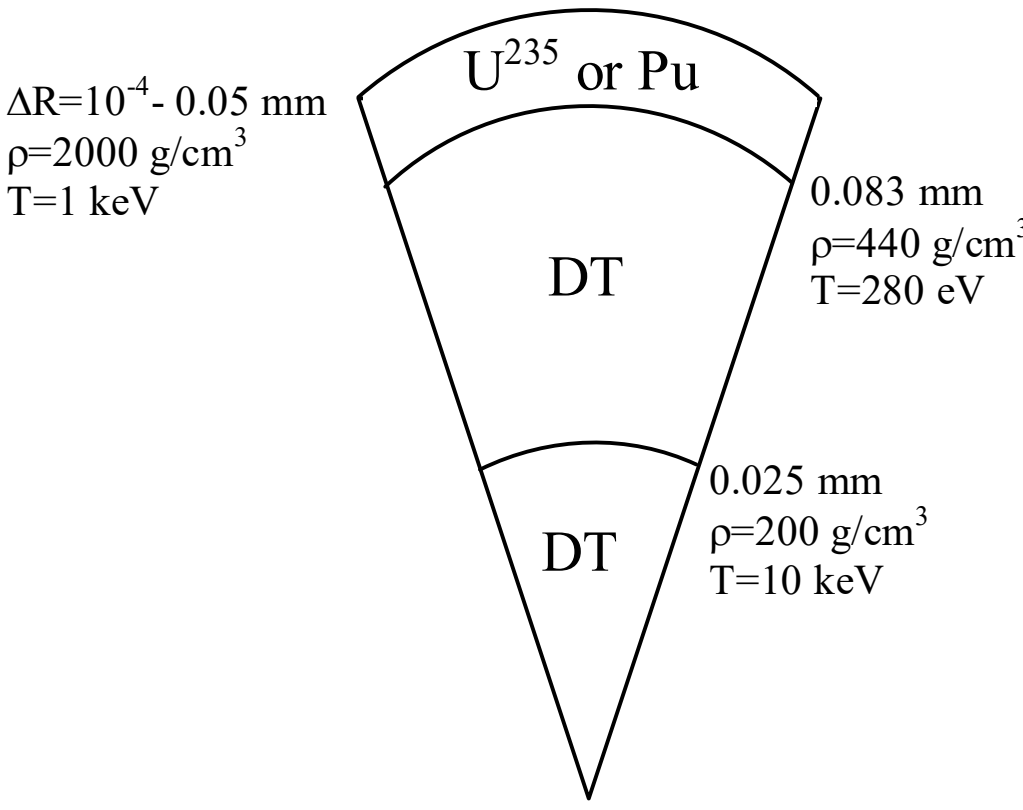

**Figure 5**: Initial conditions for calculating the performance of a deuterium-tritium target with a tamper of 93% enriched uranium or plutonium. Various thicknesses of the tamper are considered. The system is completely subcritical ($k_{eff}$=0.146, for $U^{235}$ and ΔR=0.05 mm).

The calculations were performed by a neutronic-radiation-hydrodynamic code developed by the author,[17] and to which was incorporated, as an option, the numerical solution of the neutron transport equation (4) using the characteristic method presented here. This code solves the time-dependent neutron transport equation coupled in time to radiation-hydrodynamics and to fission and fusion burnup equations.

There is no space to present in details the models used in this code for the calculations (some are explained in my book[18]), but some are enumerated below:

1. The radiation transport is performed by the flux-limited conduction approximation, with the Rosseland mean free path calculated by a semiempirical analytical expression, $\ell_R^a = aT^b/\rho^c$, for the heavy elements uranium and plutonium (the coefficients a, b and c were derived from ref.[19]); and for light compositions such as DT, LiDT or LiD (all admitted completely ionized) we use a Zeldovich-Raizer expression[20] for absorption by the inverse bremsstrahlung process. Only Thomson scattering is considered. Electron thermal conduction is accounted for and the energy of the charge particles from fusions is considered to be deposited locally.

2. The average degree of ionization for the heavy elements uranium and plutonium is calculated using an extensive phenomenological formula[21,11,18] that reproduces results from the Thomas-Fermi model[22,23] for a given density and temperature. For example, for the initial density and temperature of uranium of Figure 5 (2000 g/cm$^3$ and 1 keV) the average degree of ionization is $Z_e$=46. In these circumstances, the free electron numeric density is $n_e$= 2.33x10$^{26}$/cm$^3$ and the degeneracy temperature is $T_F$=1.38 keV (near to the assumed initial temperature). Ionization energy is also taken into account from the Thomas-Fermi model.[24,18] Due to much higher temperatures with the evolution of the system, the ideal gas model was used both in DT and uranium or plutonium plasmas (ionization energy included in the lasts).

3. The up to 10 MeV Hansen-Roach cross sections[25] were used, with an addition of an upper energy group ranging from 10 to 15 MeV (for better representing the 14 MeV neutrons emitted in D-T fusion reaction). The data from this group (including n,2n cross sections) were obtained from ref.[26][(*)]. Fission products are accounted for by a fictitious element representing them. The 10 upper energy groups (E>0.1 keV) were considered enough to represent the neutron spectra in all cases treated here (see the group structure in Fig.13), but the existence of neutron fluxes in all these groups will depend, as explained in item 6 above, on the local temperature; up to 32 angular directions can be considered and spatial discretization was the following: 30 points in the DT ignition region, 100 in the principal DT fuel and 30 in the fissile tamper (refining did not alter the results significantly).

4. Although the code solves the hydrodynamic equations in one-temperature model, a artifice (suggested by Thair[27]) was used to simulating differences in the temperatures of matter and radiation, mainly when part of the system becomes transparent to radiation. In this case, the energy of the blackbody radiation in a given mesh, $(4\sigma/c)T^4_{i-1/2}$, is reduced by a factor $f_{i-1/2}$ given by the mesh thickness divided by the Planck mean free path in the mesh, with the limits $f_{i-1/2}=\{1,10^{-3})$. It is as if $T_r=(f)^{1/4}T$, where T is the matter temperature (of ions and electrons). (In the calculations, the values of f have became lower than 1 only in transparent regions of the DT plasma, since the uranium and plutonium plasmas are extremely opaques, even in extreme temperatures, emitting as blackbody radiation.) All other radiation variables (as the radiation pressure) are affected by the parameter f.

Figure 6 and Table 1 show the results of the microexplosion for several thicknesses of the enriched uranium and plutonium tampers. In Table 1 we also show the percentage contribution of each region to the energy liberated in the case of uranium tamper, as well as the thermal radiation energy that leaked out from the target and the energy deposited by neutrons (all the results for the plutonium are similar to those of enriched uranium). Note that extrapolating the energy to zero thickness we obtain an energy of ≅ 120 MJ, near the

---

[(*)] As we do not have T, $He^4$ and $He^3$ cross sections, they are taken equal to those of deuterium. The impact of this on the results was estimated to be very low, partly because, like for deuterium, their absorption cross sections are practically nulls in all energy range of interest ((n,p) $He^3$ are high).[26]

energies presented in ref.[4] for this target pellet for many options admitted in the MEDUSA-IB[28] code used in these calculations.

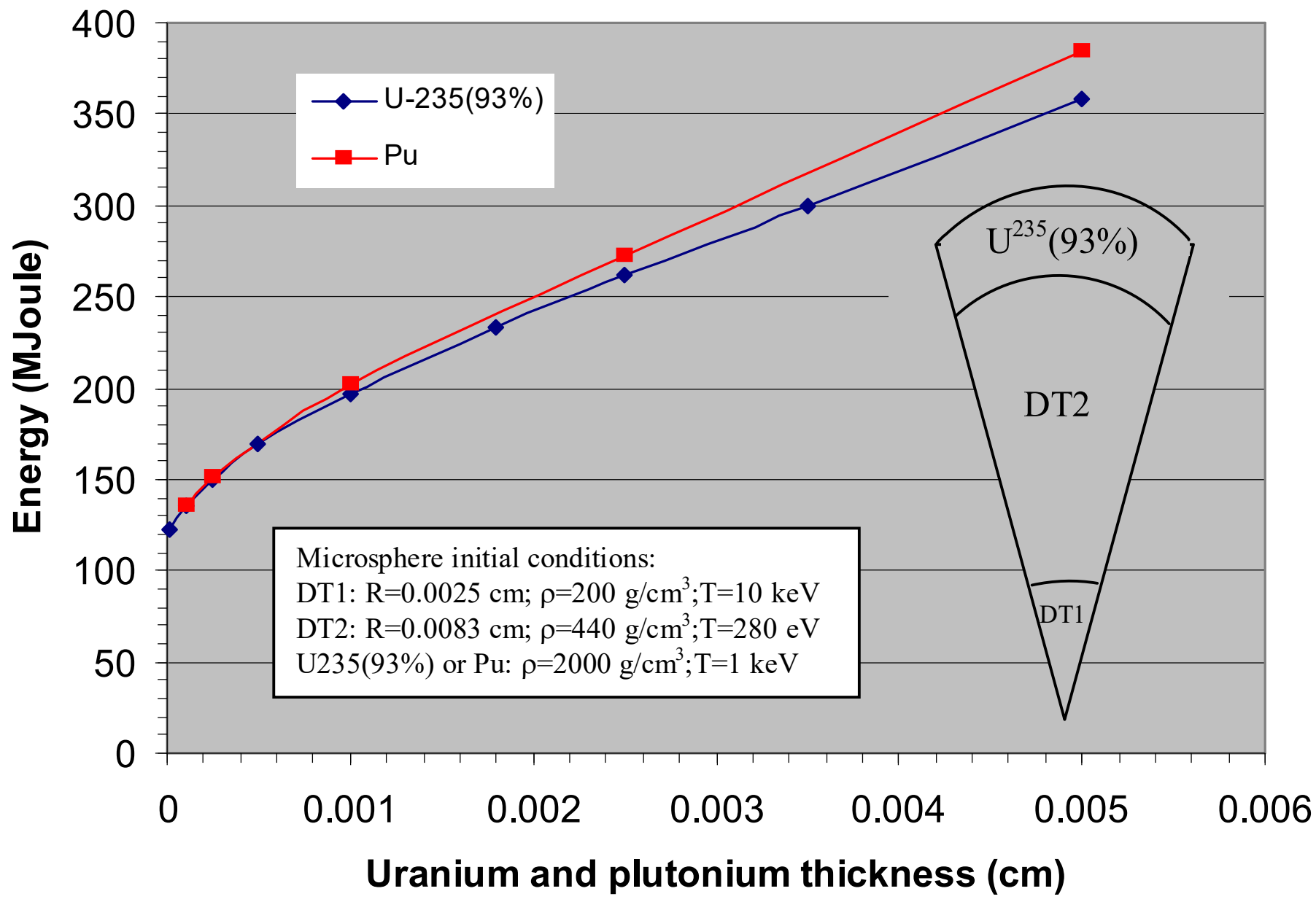

**Figure 6**: Energy yielded by the microexplosion of the deuterium-tritium microsphere specified in Fig. 5 for several thickness of the tamper composed by 93% enriched uranium or plutonium.

As can be seen, these results show a significant increase in microexplosion energy with the addition of a uranium or plutonium tamper. Besides the benefits of using a heavy element as a tamper in the DT targets (well explained by Fraley et al.[29]), there is a much more and obvious advantage in using fissile elements in this tamper, because fission contribution is very significant to the total energy yield. Fusion is also enhanced by neutrons produced by the fission reactions (and vice-versa) and by the hydrodynamics consequences of the fission explosion in the tamper (that does not react only passively as a non fissile material does).

Figure 7 shows the time evolution of the total energy produced by the micro-explosion, the portion due to fusion, the radiation leaked out from the system and the energy deposited by neutrons for uranium thickness of 0.005 cm. Note that the leakage of thermal radiation increases with the uranium thickness (Fig.8), because uranium emits as blackbody radiation and more radiation is emitted as more mass of uranium exist.

In Figure 9 we have the temperature distributions at four successive times. These distributions reflect the progress of a thermonuclear wave from the ignition region to the main thermonuclear fuel.

**Table 1**: Total energy produced by the microexplosion of the deuterium-tritiun-$U^{235}$ target pellet (Fig.5) as a function of thickness of the $U^{235}$(93%) tamper. The table shows also the percentage contribution of each region to the energy produced, the energy of thermal radiation leaked out from the pellet and the energy deposited by neutrons downscattering (mainly in the DT moderator regions); 1 MJ=$10^{13}$ ergs.

| $\Delta R^U$ (cm) | $U^M$ ($10^{-3}$g) | E (MJ) | DT1 (%) | DT2 (%) | U-235 (%) | Leakage of thermal radiation (MJ) | Neutron energy deposited (MJ) |
|---|---|---|---|---|---|---|---|
| 0.00001 | 0.017 | 123,9 | 2,12 | 97.84 | 0.04 | 3.18 | 26.0 |
| 0.0001 | 0.17 | 135.5 | 2.03 | 97.3 | 0.68 | 3.75 | 28.4 |
| 0.00025 | 0.45 | 150.1 | 1.92 | 96.1 | 1.99 | 6.69 | 31.7 |
| 0.0005 | 0.92 | 168.9 | 1.76 | 94.0 | 4.27 | 11.1 | 35.1 |
| 0.001 | 1.95 | 197.2 | 1.51 | 89.7 | 8.79 | 18.9 | 39.8 |
| 0.0018 | 3.84 | 233.1 | 1.27 | 83.2 | 15.5 | 31.0 | 44.0 |
| 0.0025 | 5.76 | 261.2 | 1.13 | 78.0 | 20.9 | 42.0 | 46.5 |
| 0.0035 | 8.97 | 300.0 | 0.98 | 71.3 | 27.7 | 58.8 | 49.2 |
| 0.005 | 14.9 | 357.8 | 0.84 | 62.7 | 36.5 | 85.8 | 51.9 |

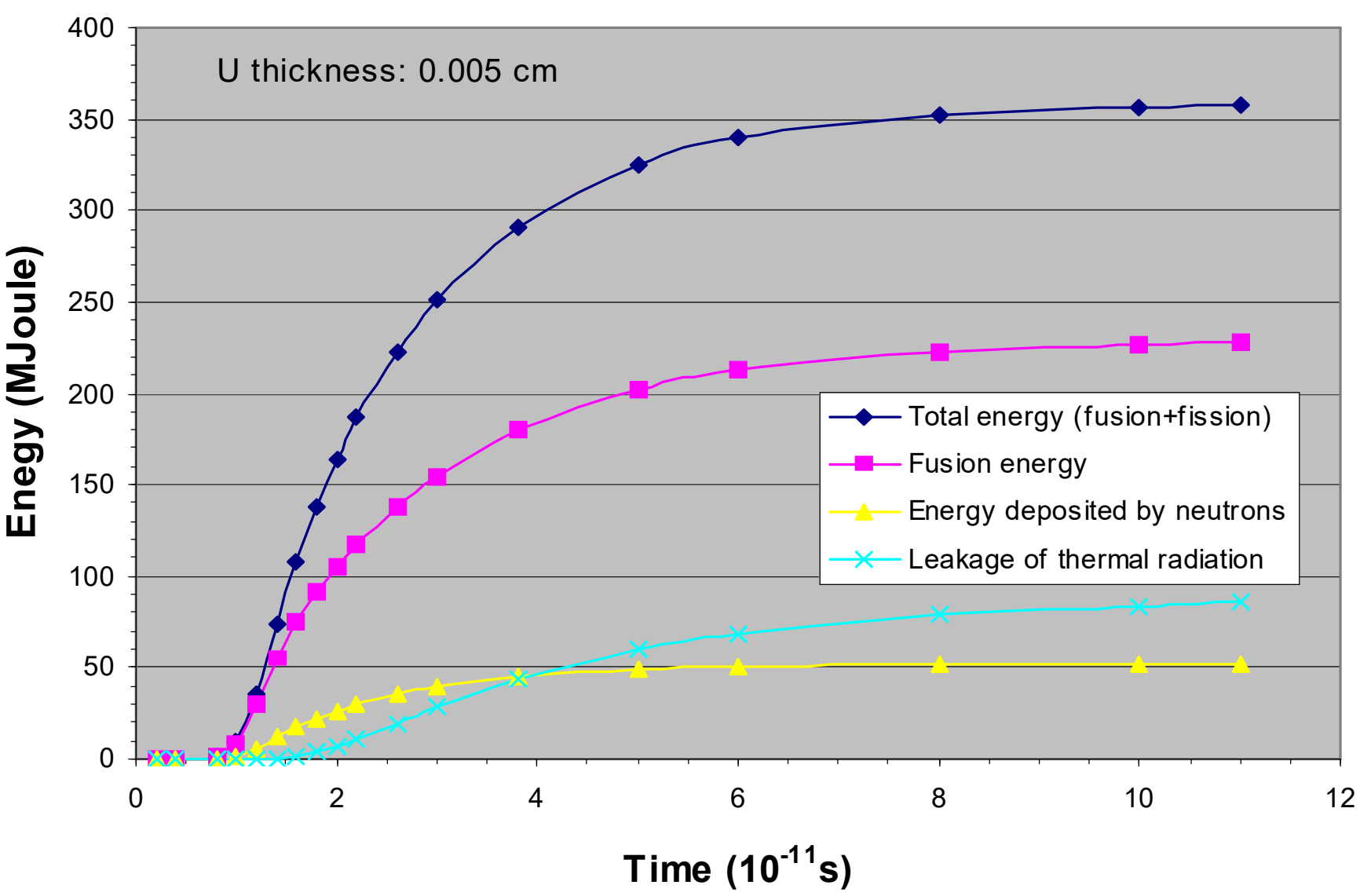

**Figure 7**: Time evolution of energy yielded by the pellet for uranium thickness of 0.005 cm.

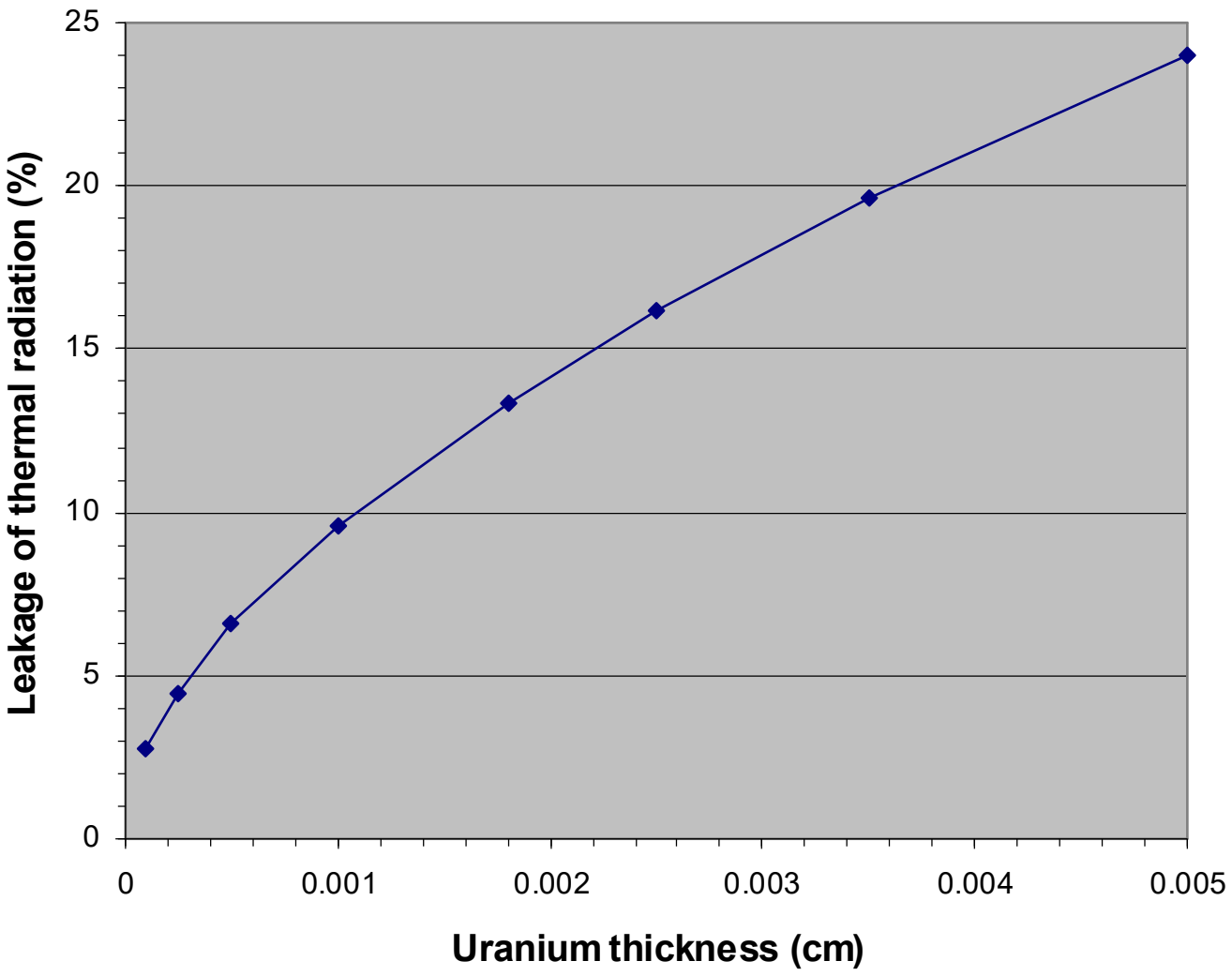

**Figure 8**: Percentage leakage of thermal radiation energy as a fuction of uranium thickness. Extrapolating the curve to zero thickness, we obtain a value near to that calculated in ref.[4].

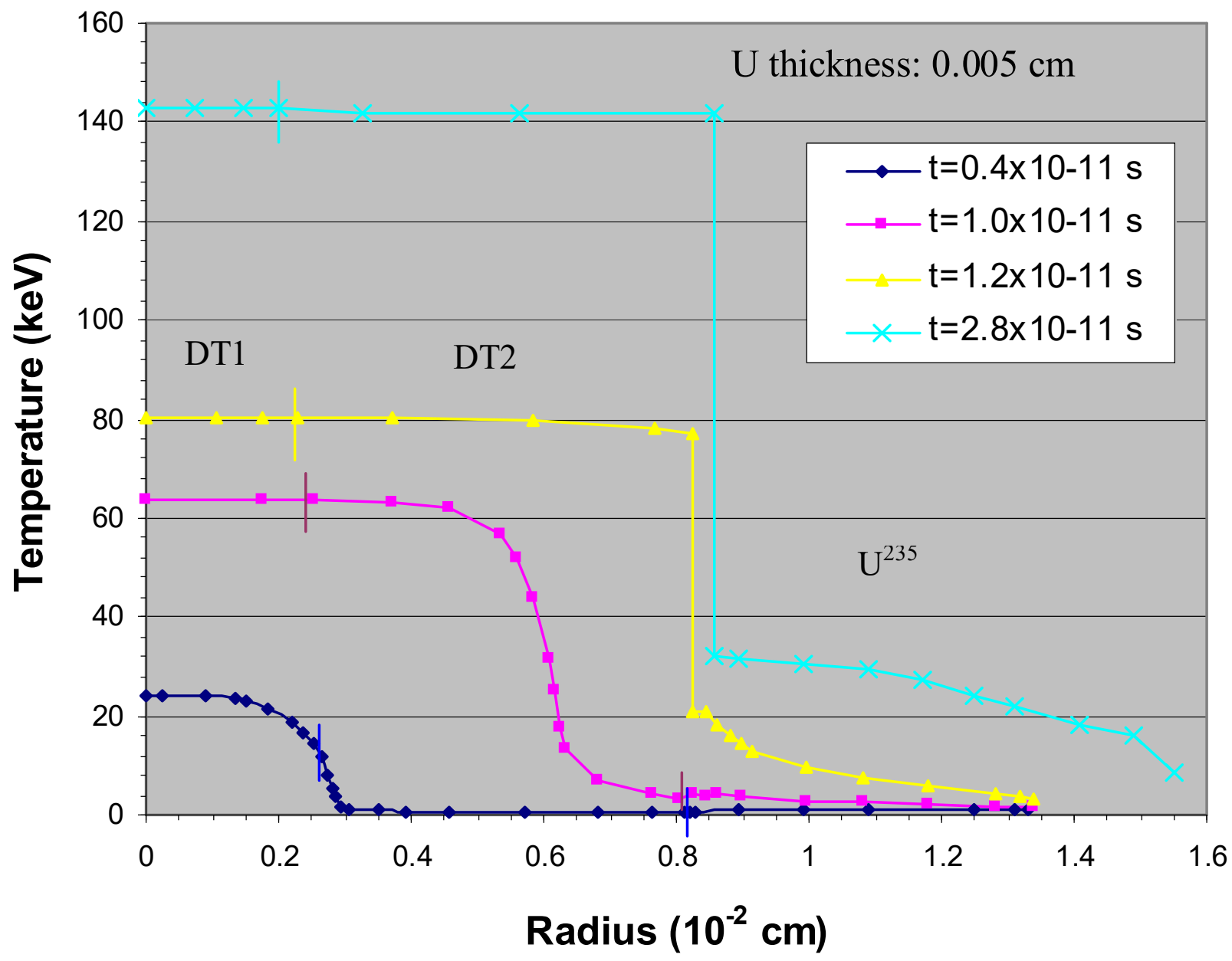

**Figure 9**: Time evolution of the spatial distribution of temperature for uranium thickness of 0.005 cm. The frontiers of the regions are marked with vertical bars.

The progress of the DT burnup and the final burnup of the DT and enriched uranium are shown in Figures 10 and 11, respectively. The higher burnup of the internal region of the principal DT fuel in the two middle curves of Figure 10 is due to higher density of this fuel, that also undergo a much great influence of the tamper region. Note that the first DT region is slightly re-imploded by the energy produced in the main fuel.

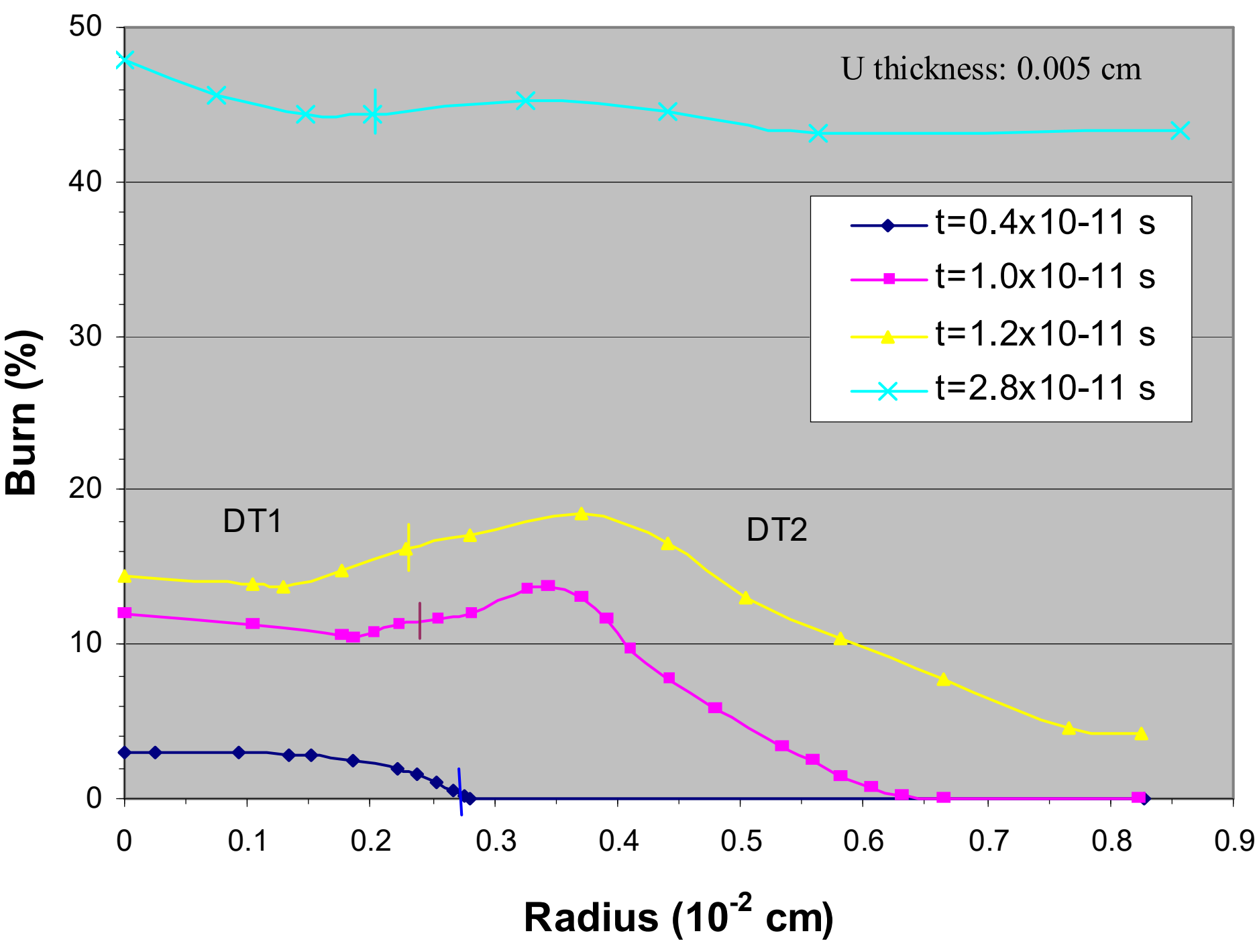

**Figure 10**: Spatial distribution of deuterium burning at four different times and for uranium thickness of 0.005 cm. (See Fig.9 for the progress of thermonuclear wave.)

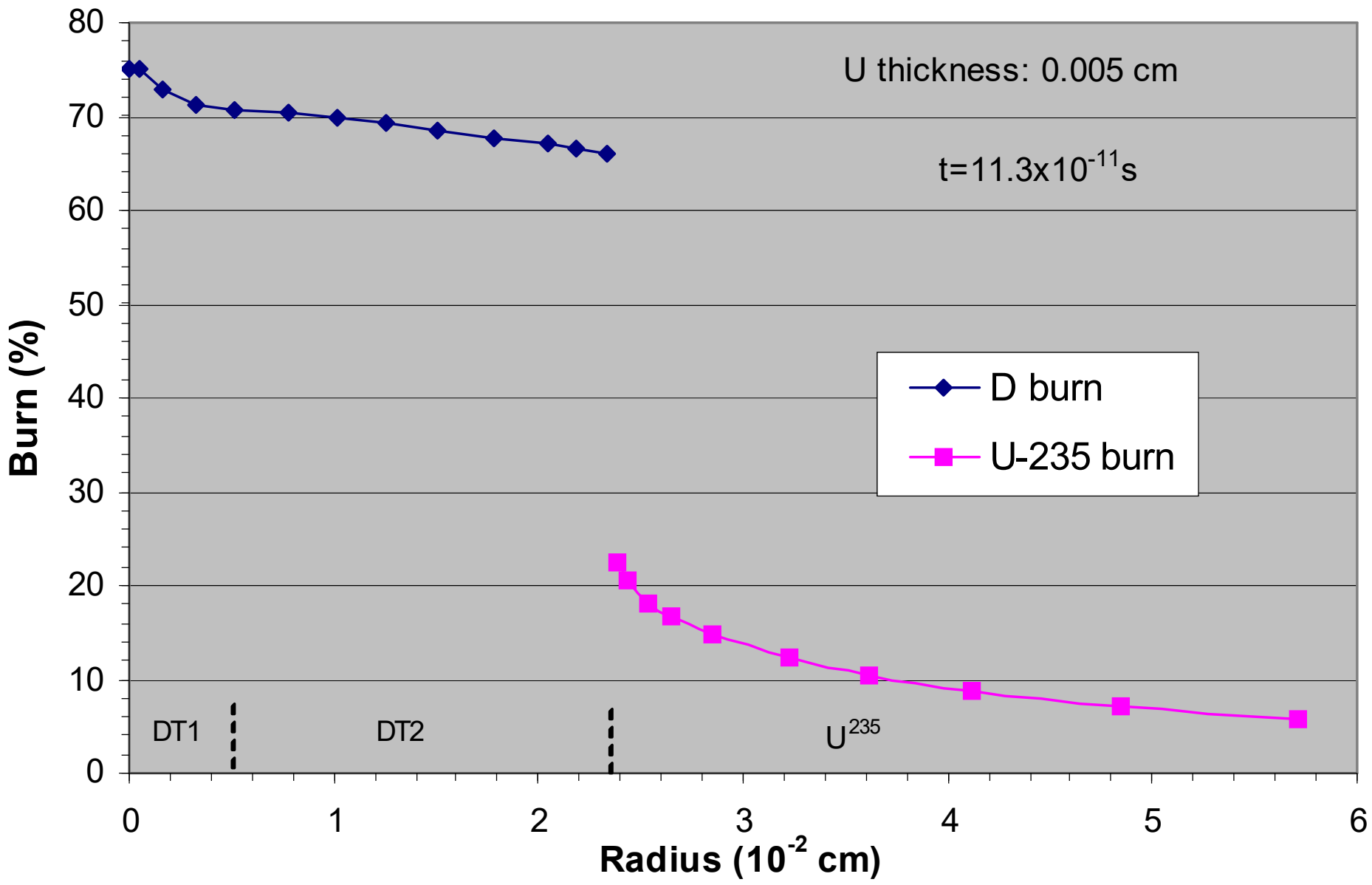

**Figure 11**: Final burnup of deuterium and uranium-235.

The influence of the uranium enrichment on the energy produced by the fusion-fission microexplosion is seen in Table 2. This table shows that even with lower enriched uranium the contribution of fission to the energy is meaningful, because of the presence of the $U^{238}$ whose fast fissions in hard neutron spectra are very great.

**Table 2**: Influence of the uranium enrichment on the results for 0.005 cm uranium thickness. The tamper contribution is the percentage contribution to the total energy produced.

| $\varepsilon^{235}$(%) | E (MJ) | Tamper contribution (%) | Burn by fission of $U^{235}$ and $U^{238}$ (%) | Burn by fission of $U^{238}$ (%) |
|---|---|---|---|---|
| 0.7 | 293 | 22.4 | 5.54 | 5.29 |
| 3.0 | 294 | 22.8 | 5.66 | 5.15 |
| 10 | 298 | 23.8 | 6.00 | 4.78 |
| 25 | 308 | 26.4 | 6.86 | 3.98 |
| 50 | 325 | 30.4 | 8.35 | 2.65 |
| 70 | 339 | 33.4 | 9.56 | 1.59 |
| 93 | 358 | 36.5 | 11.0 | 0.37 |

Figure 12 displays the spatial distribution of neutron flux at three consecutive times and Figure 13 the neutron flux by group in four different locations and at time $t=10^{-11}$s. In the early stages of the thermonuclear microexplosion the neutron spectra are very fast because there is a predominance of the D-T fusion reactions over the fission reactions in the tamper.

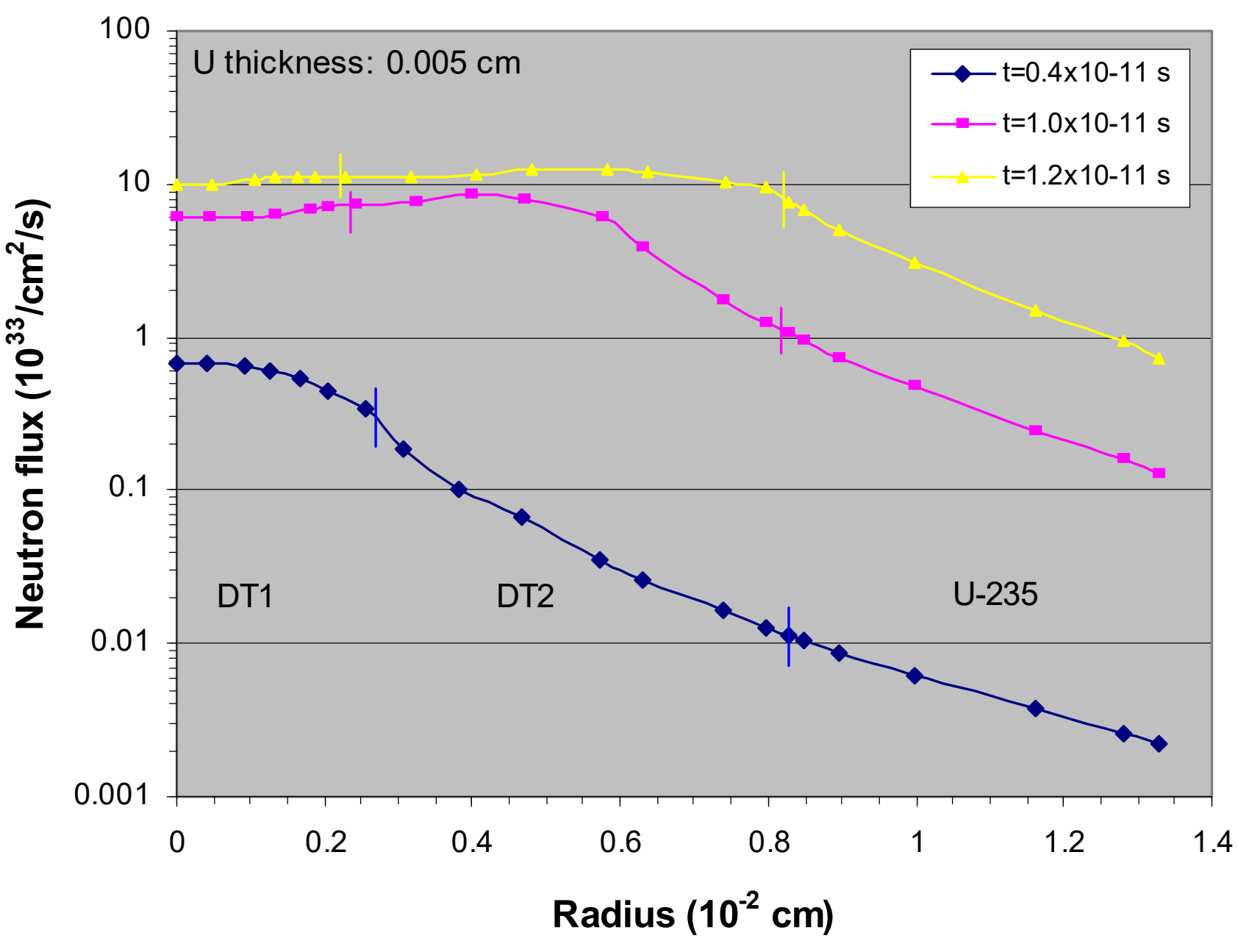

**Figure 12**: Spatial distribution of neutron flux (summed over all groups) at three different times. The vertical bars separate the regions at each time.

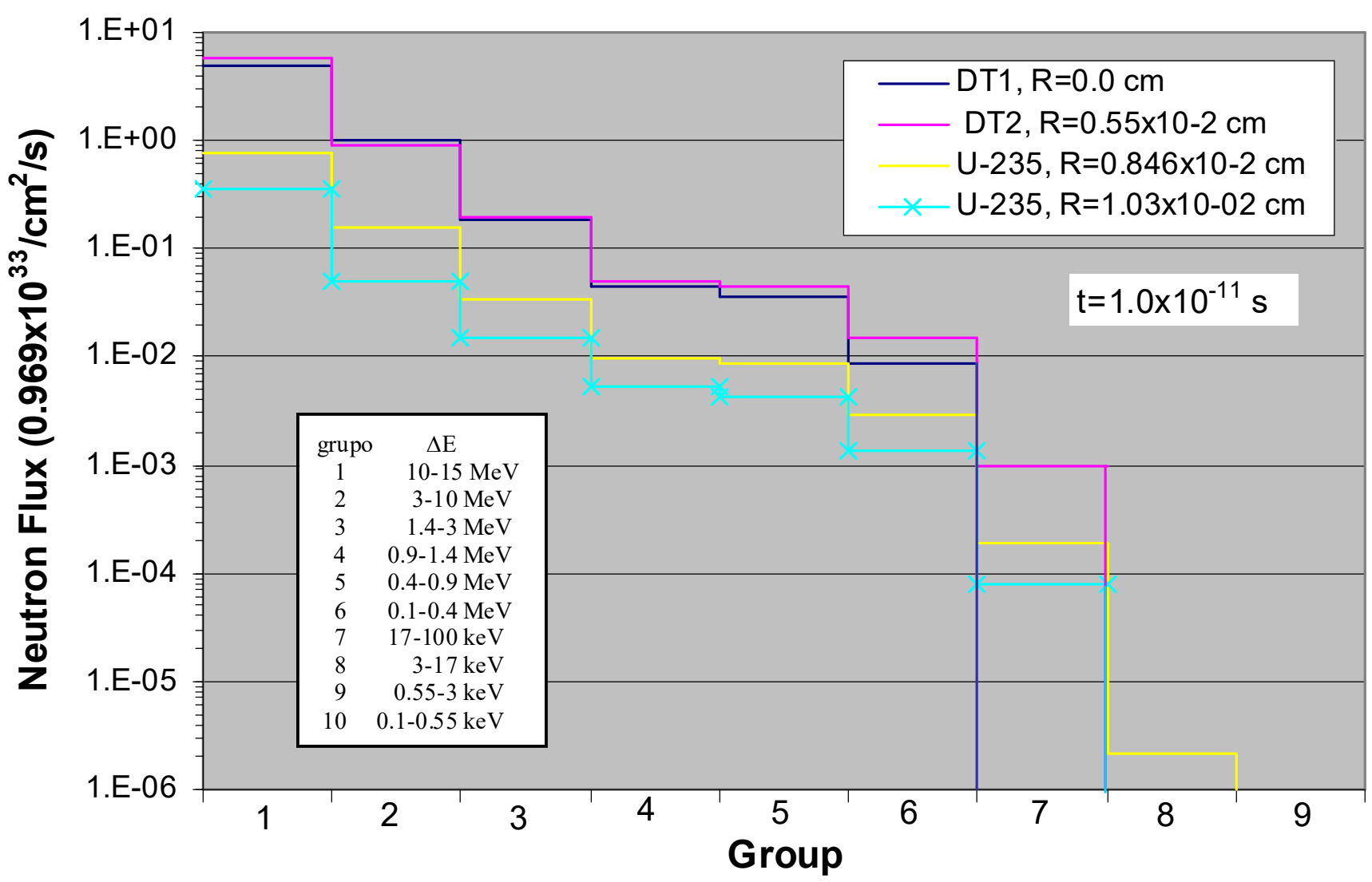

**Figure 13**: Neutron flux (in units of $0.969\text{x}10^{33}$ neutrons/cm$^2$/s) by energy groups in four different locations and at time t=$1.0\text{x}10^{-11}$s; (identify the locations in Fig.12 for the corresponding time). The energy intervals of the groups are shown. D-T and D-D fusion neutrons born in groups 1 and 3, respectively

The behavior of the time-dependent neutron transport solution was very good in all stage of the calculations, as was proved by the neutron balance executed by the code in all time-steps. The error in neutron balance (Fig.14) remains lower than 2% during most of the calculations.

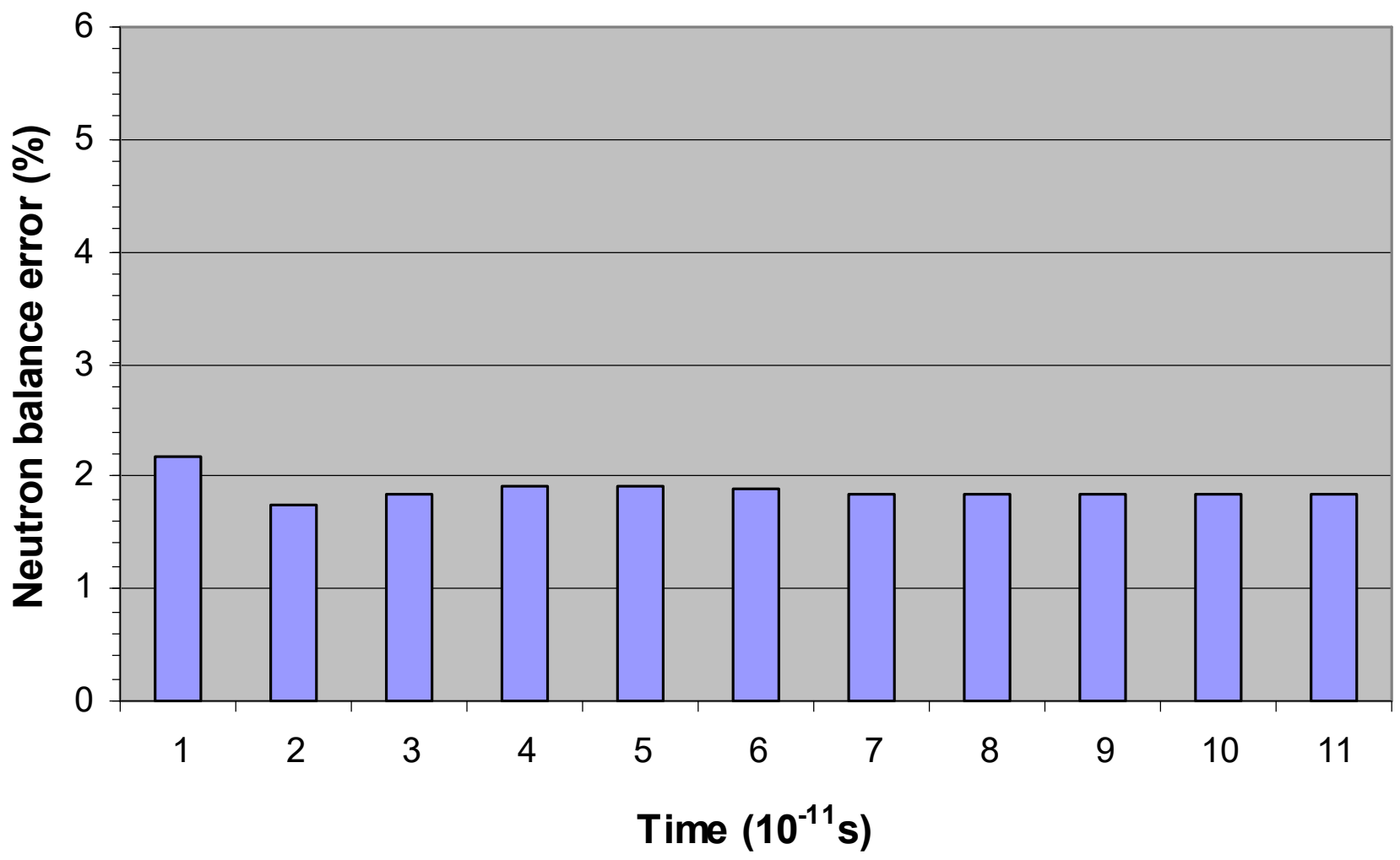

**Figure 14**: Percentage error in neutron balance performed in various stages of the time-dependent transport solution using the characteristic method.

Below, we present the final inventory of the isotopes in all regions and a typical balance of neutrons performed by our code (at time t=11.2x$10^{-11}$s and for $\Delta R^U$=0.005 cm). The mixtures 5, 6 and 7 correspond, respectively, to the uranium region (we disregarded the $U^{236}$ production), to the DT ignition region and to the main DT fuel of the target pellet.

```
MIXTURE NUMBER:  5
REMAINDER MASS FRACTION OF U-235/U-236 ISOTOPES RELATIVE TO THE TOTAL MASS OF THE MIXTURE AFTER DEPLETION
U-235:0.822339D+00 U-236:0.000000D+00
FRACTION OF U-238 BURNED BY FISSIONS:0.542405D-01

MIXTURE NUMBER:  6
REMAINDER ATOMIC FRACTION OF DEUTERIUM IN THE MIXTURE:0.143952D+00
REMAINDER ATOMIC FRACTION OF TRITIUM IN THE MIXTURE:0.168061D+00
FRACTION OF D BURNED BY FUSIONS:0.712095D+00
TOTAL NUMBER OF FUSIONS: 0.109319D+19
FRACTION OF D-D FUSION REACTION TO THE TOTAL NUMBER OF FUSIONS: 0.271290D-01
FRACTION OF D-T FUSION REACTION TO THE TOTAL NUMBER OF FUSIONS: 0.971144D+00
FRACTION OF D-HE3 FUSION REACTION TO THE TOTAL NUMBER OF FUSIONS: 0.172705D-02

MIXTURE NUMBER:  7
REMAINDER ATOMIC FRACTION OF DEUTERIUM IN THE MIXTURE:0.160438D+00
REMAINDER ATOMIC FRACTION OF TRITIUM IN THE MIXTURE:0.184608D+00
FRACTION OF D BURNED BY FUSIONS:0.679124D+00
TOTAL NUMBER OF FUSIONS: 0.815245D+20
FRACTION OF D-D FUSION REACTION TO THE TOTAL NUMBER OF FUSIONS: 0.285997D-01
FRACTION OF D-T FUSION REACTION TO THE TOTAL NUMBER OF FUSIONS: 0.969684D+00
FRACTION OF D-HE3 FUSION REACTION TO THE TOTAL NUMBER OF FUSIONS: 0.171582D-02

  BALANCE OF NEUTRONS
  TOTAL NUMBER OF NEUTRONS CURRENTLY PRESENT IN THE SYSTEM=  0.1248D+19
  TOTAL NUMBER OF NEUTRONS LEAKED OUT FROM THE SYSTEM=  0.9995D+20
  TOTAL NUMBER OF NEUTRONS ABSORBED (CAPTURE+FISSION)=  0.4257D+19
  TOTAL NUMBER OF NEUTRONS PRODUCED BY ALL FISSIONS=  0.1605D+20
  TOTAL NUMBER OF NEUTRONS PRODUCED BY ALL FUSIONS=  0.8117D+20
  TOTAL NUMBER OF NEUTRONS PRODUCED BY ALL (N,2N) REACTIONS=  0.6283D+19
  PERCENTAGE ERROR=100xABS{1-[N(INITIAL)+N(PRODUCED)]/[N(SYSTEM)+N(ABSORBED)+N(LEAKAGE)]}=  0.1844D+01(%)
```

By way of illustration, 56,952 time-steps or hydrodynamics cycles were performed in the calculations for uranium thickness of 0.005 cm. For 16 angular directions, 10 energy groups and 130 spatial meshes, the number of simultaneous equations solved only in neutronic calculations was 56,952x16x10x130=1.185 billions. Even so, the CPU time of global calculations (including radiation-hydrodynamic, burnup etc.) was only 5.9 minutes in a present day microcomputer! (How long would it take, for example, in 1970-80 decades?)

### 7.2 A fission-fusion microexplosion in a complex target with Li composition in the thermonuclear fuel.

The target considered in this item is shown in Figure 15. In this case, we have a microexplosion of a supercritical fissile mass of plutonium (covered by a tiny layer of beryllium), whose objective is to ignite a thermonuclear $Li_{0.5}D_{0.5}$ (90% $Li^6$ enriched) fuel in a superior adjacent region covered by a layer of natural uranium tamper. The initial thermonuclear reactions in the internal mass of $Li_{0.5}D_{0.25}T_{0.25}$ (natural Li) are intent to ensure a very potent neutron source necessary to generate a persistent fission chain reaction into the fissile mass. The isotopic composition of (weapon grade) plutonium was: $Pu^{239}$(93.8%), $Pu^{240}$(5.8%), $Pu^{241}$(0.35%) and $Pu^{242}$(0.05%). The initial criticality of the system is $k_{eff}$=1.89 (the time eigenvalue is $\alpha$=3.67x10$^{10}$/s). We used 32 angular directions and 10 energy groups (with "upscattering" explained in item 6); spatial discretizations were: 30 points in LiDT region, 100 in the plutonium, 30 in the beryllium, 70 in the LiD and 30 in the natural uranium.

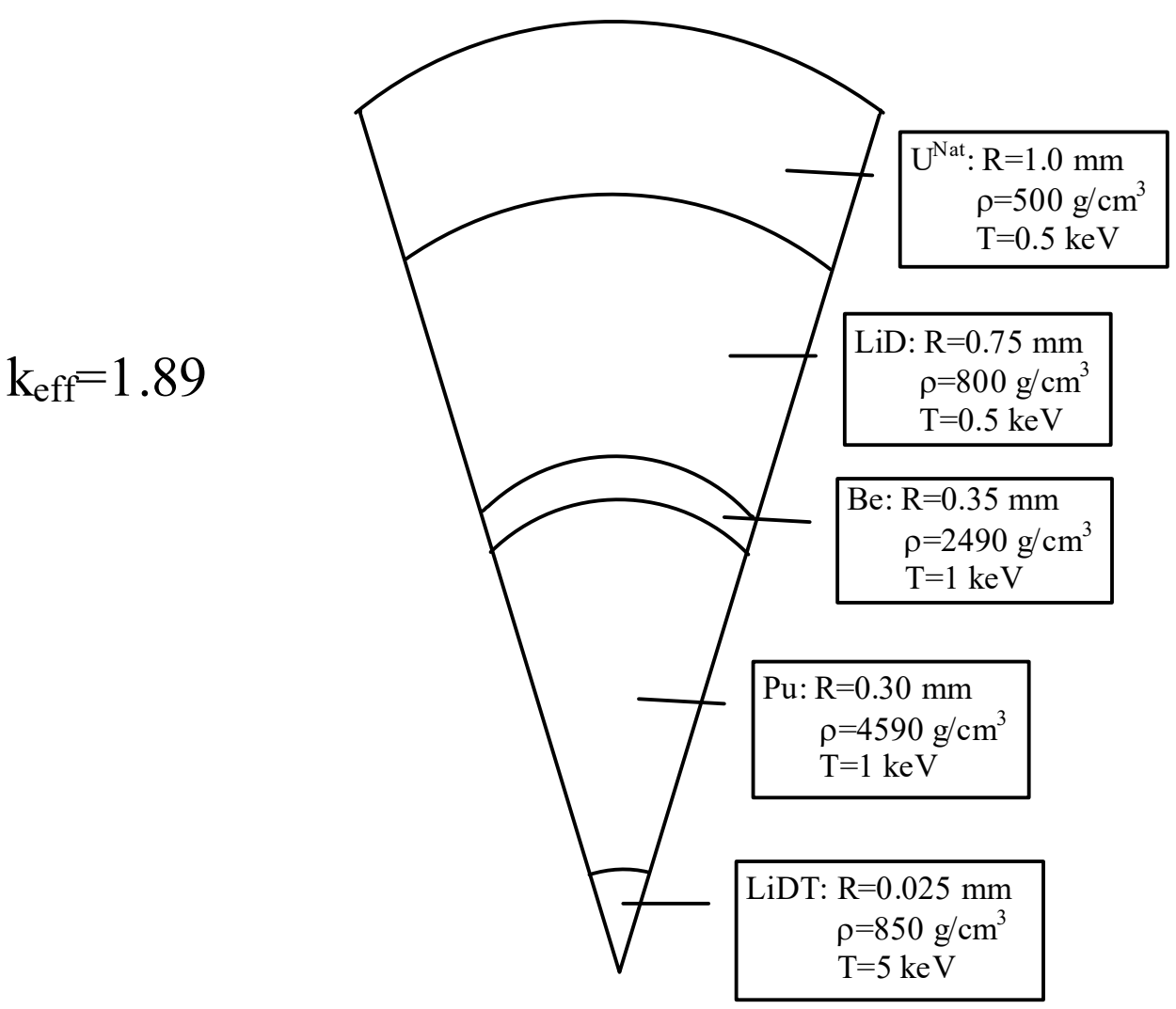

**Figure 15**: Multilayer target considered in the second case. In the $Li_{0.5}D_{0.25}T_{0.25}$ region, we considered the natural isotopic composition of Li: $Li^6$(7.42%)+ $Li^7$(92.58%); in the $Li_{0.5}D_{0.5}$ region, we took it 90% $Li^6$ enriched. Initially, $k_{eff}$=1.89.

This much more complex system resembles the process of a thermonuclear explosion in the secondary module of a modern thermonuclear bomb.[18]

As in the previous case, we do not discuss here the feasibility of the initial conditions for this target pellet, which are completely arbitrary (the plutonium and beryllium densities were the same as adopted in ref.[30]). Our main interest is, as before, to test the consistence and stability of the neutronic solution. We do not discuss also in details the performance of the target.

Figure 16 shows that the total energy produced by the fission-fusion microexplosion was about 25 ton of TNT, with 25% coming from fissions in the plutonium mass, 70% from fusions in the LiD main fuel region (including the energy produced by the 4.8 MeV ($n,Li^6$) reactions) and 5% from fissions in natural uranium region (the energy from the internal LiDT region was negligible). Static $k_{eff}$ calculations show the fall of criticality of the whole target due to burnup and expansion of the plutonium.

Figure 17 presents the evolution of D and $Li^6$ burn in the main fuel, the last by the tritium production reaction $n(\alpha,T)Li^6$. Percentage contributions of the three main fusion reactions to the D burn are also shown. The results in this figure can be explained in the following way. Initially, when there is no T in the main fuel, only D-D reactions occur (a few reactions in this time). Next, with the production of T by one branch of D-D reactions and mainly by ($n,Li^6$) reactions (neutrons coming mainly from the fissions in the plutonium), the D-T reactions increase because the probability of this reactions is order of magnitude greater then the D-D reactions, in spite of much greater amount of D in the fuel. With the growth of thermonuclear reactions due to increase of the temperature in the fuel (that reaches values greater than 100 keV = $1.16x10^9$ K!), there is an equilibration between these two types of fusion reactions: D-D reactions are high because of greater quantity of D, and D-T reactions are high because of continuous production of T in the fuel and because of their much higher probabilities. (In the LiDT ignition region, D-D, D-T and D-$He^3$ reactions contributed with 2.1%, 97.74% and 0.16%, respectively.)

Figures 18 and 19 show the spatial distributions of the neutron flux and the burnup of deuterium and plutonium at two times. Note, in Figure 18, that the much high flux in the internal region of the main LiD fuel is consequence of the intense thermonuclear reactions that begin in this region and induced by the fission explosion in the inner plutonium region. The energy is transferred from the plutonium to the LiD by hydrodynamic compression, electron thermal conduction, neutrons and thermal radiation. Pre-production of tritium by the ($n,Li^6$) reactions (neutrons coming from the fission explosion) has an important influence on the thermonuclear ignition.

The final burnup of the deuterium (in the ignition region and in the main thermonuclear fuel) and the plutonium is plotted in Figure 20.

The errors in neutron balance are presented in Figure 21. They fell below 1% during most of the calculations.

Table 3 shows that the same results were obtained for two different angular discretizations of neutron flux and that close results were also obtained by not considering “upscattering” (item 6) for 10 groups (E>0.1 keV, see Fig.13) and 7 groups (E>17 keV), meaning that upper energy groups of Hansen-Roach cross sections (added by a 10-15 MeV group) have a much greater influence on the results.

For 32 angular discretizations, about 2.3 billions of simultaneous equations were solved in neutronic calculations for 27,327 time-steps performed. The global calculations, in my microcomputer, took 10.6 minutes of CPU time. At final time $t=68\times10^{-11}$s, about 9.4% of the yielded energy escaped as thermal radiation, 18.2% is carried as internal energy and 62.5% as kinetic energy (10% of the fission energy was considered not deposited in the pellet and part of energy was carried by neutrons that leaked out). The uranium external radius reaches a velocity of 8,600 km/s!

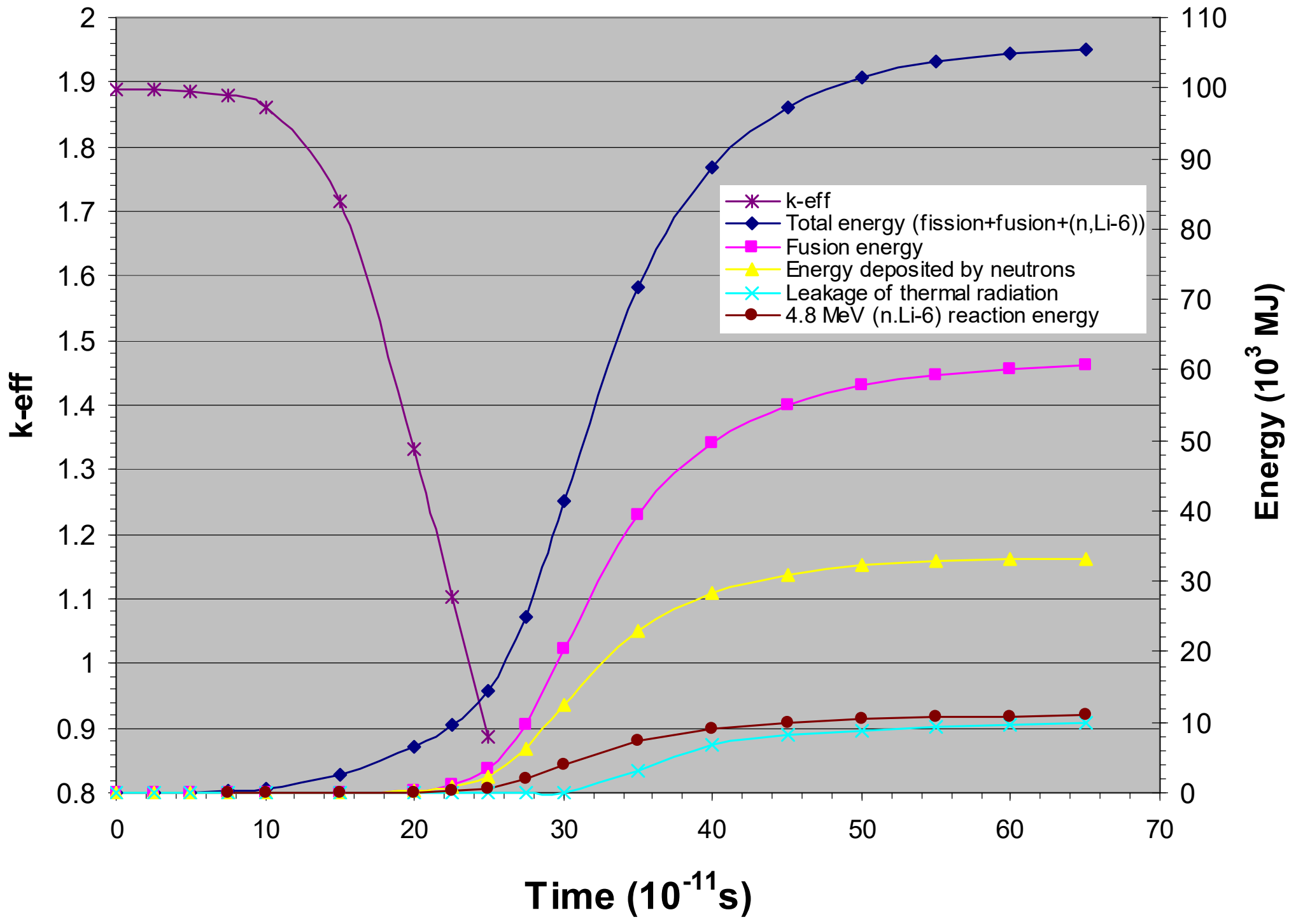

**Figure 16**: Energy yielded by the microexplosion of the target pellet in Fig.15, the contributions of fusion and (n,$Li^6$) reactions, the energy deposited by neutrons and the energy leaked out by thermal radiation. The concomitant static $k_{eff}$ calculations give the fall of criticality with the expansion and burn of plutonium. $10^3$MJ=239 kg of TNT.

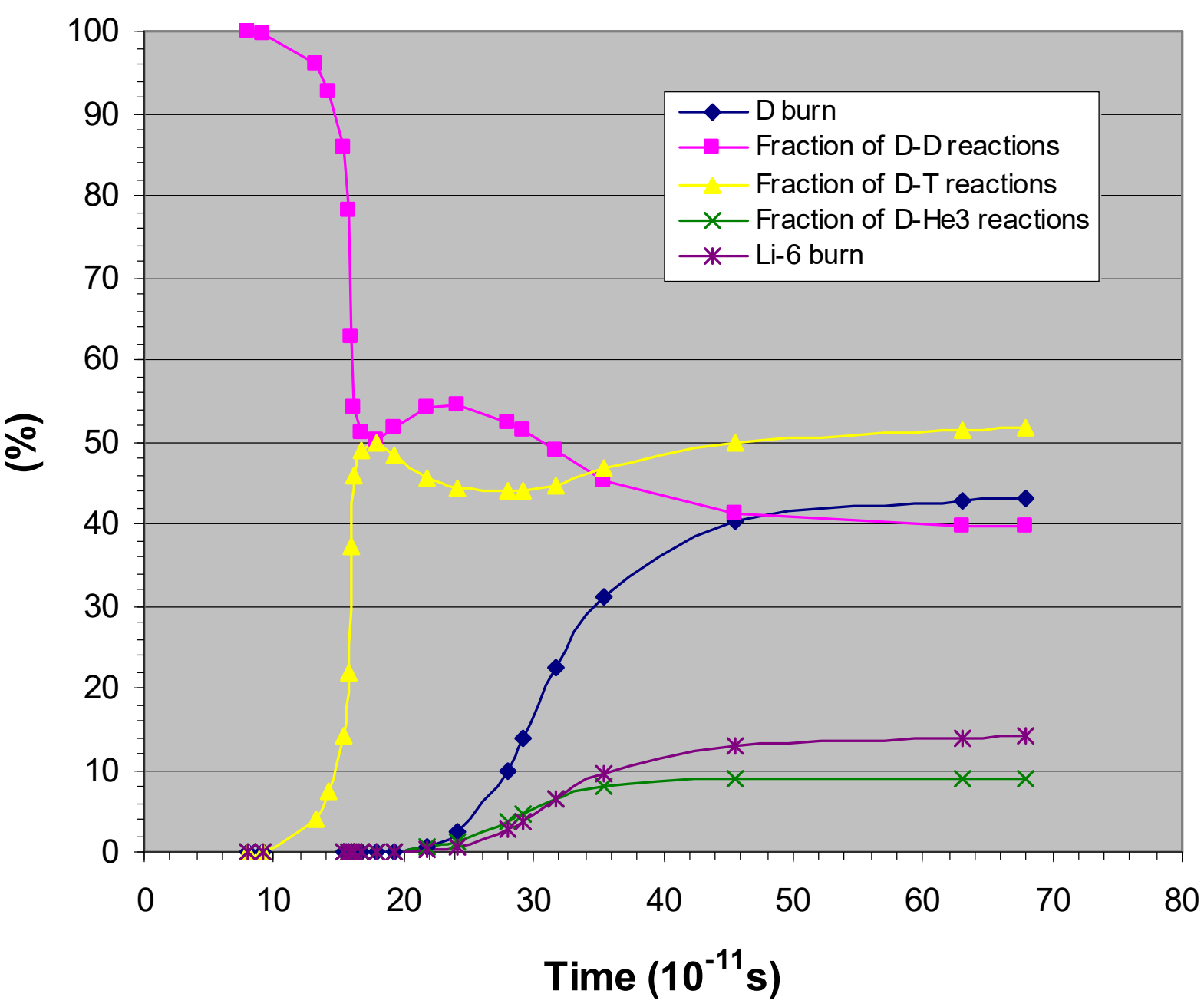

**Figure 17**: Percentage fraction of deuterium burn in the LiD main thermonuclear fuel and contributions of D-D, D-T and D-$He^3$ fusion reactions to the total number of fusions (note that in D-D reactions 2 D are destroyed); $Li^6$ burn by the tritium production n($\alpha$,T)$Li^6$ reaction is also shown.

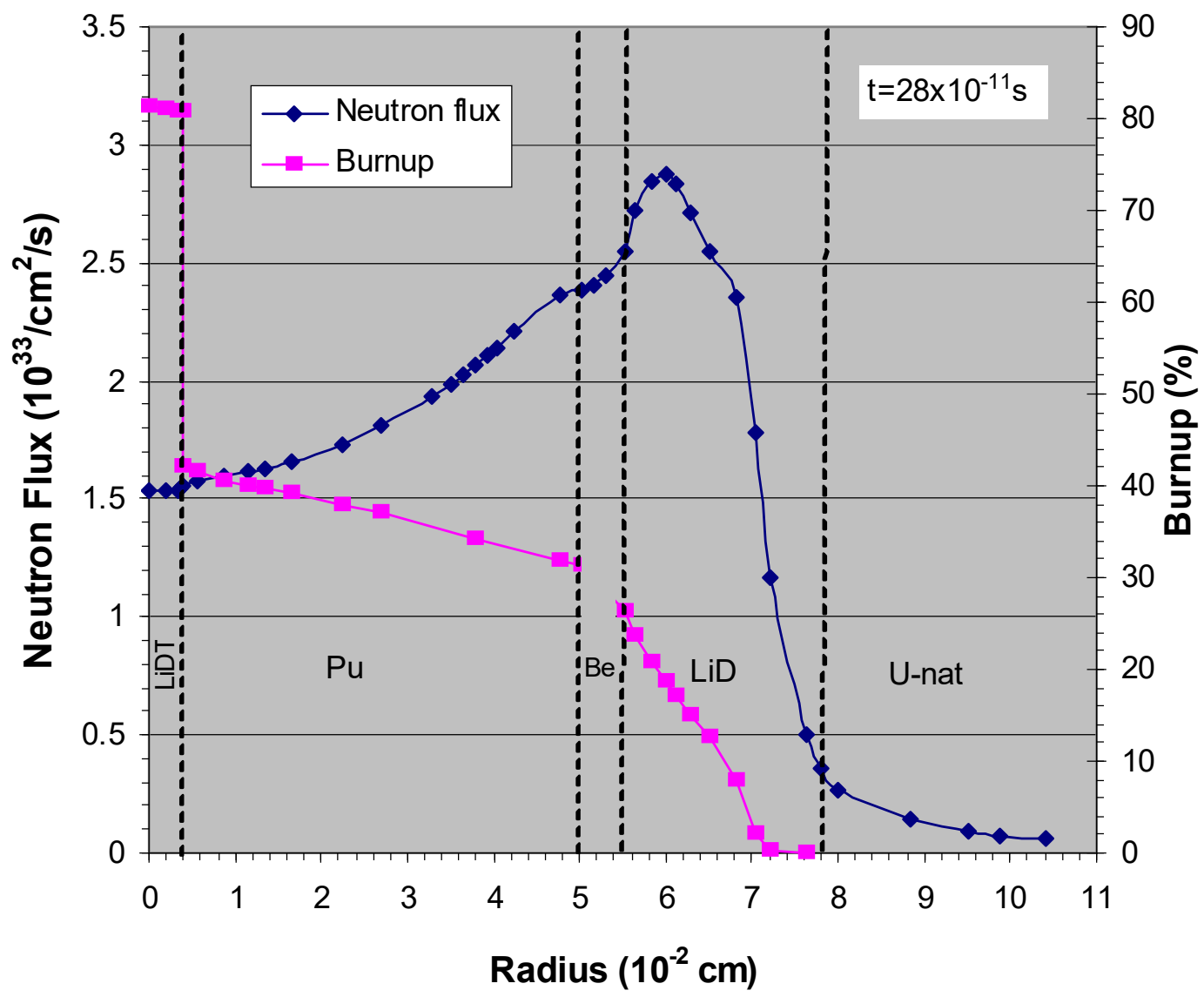

**Figure 18**: Total neutron flux (summed over all groups) and burnup of deuterium and plutonium at t=28.0x10$^{-11}$s.

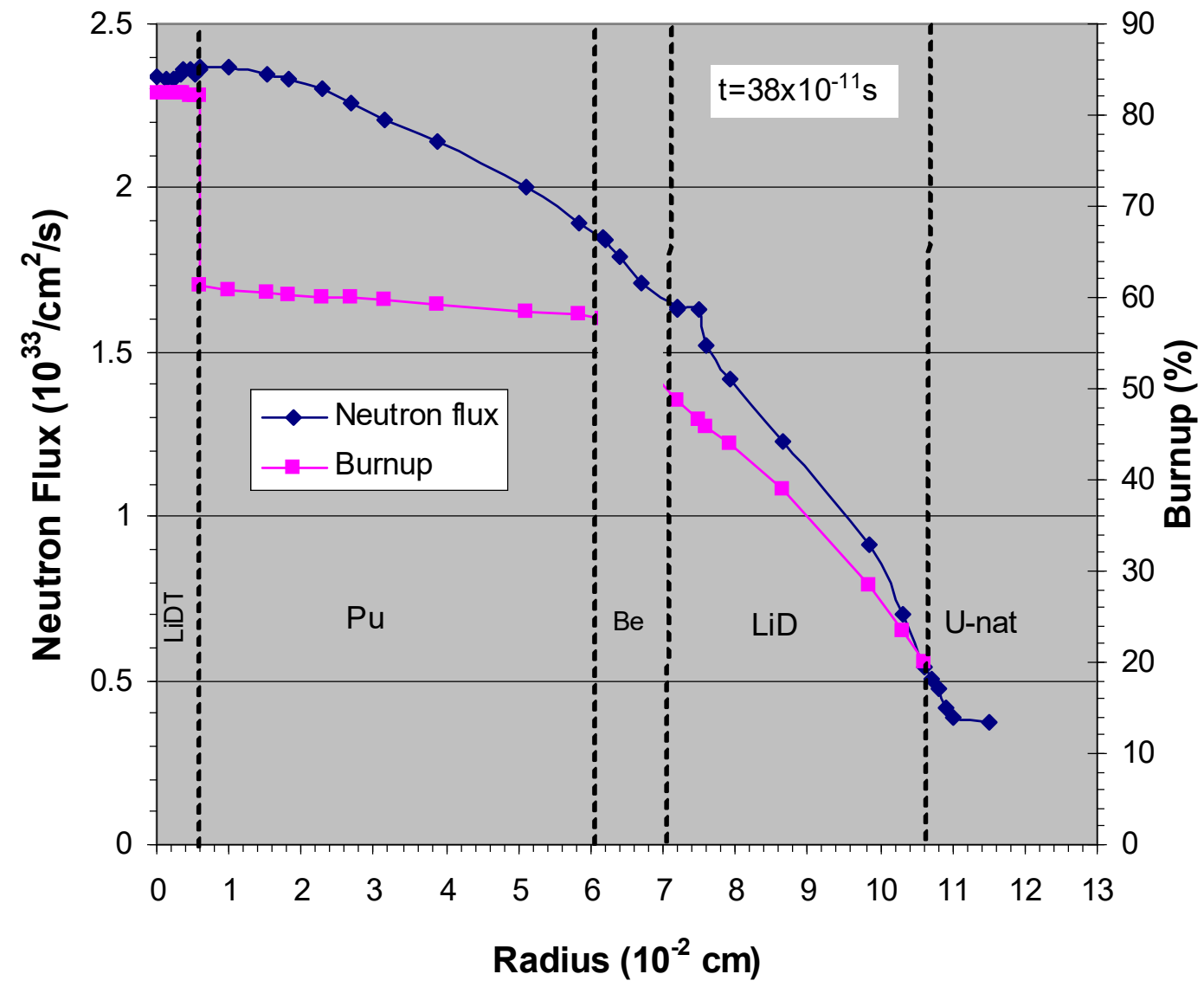

**Figure 19**: Total neutron flux and burnup of deuterium and plutonium at t=38.0x$10^{-11}$s.

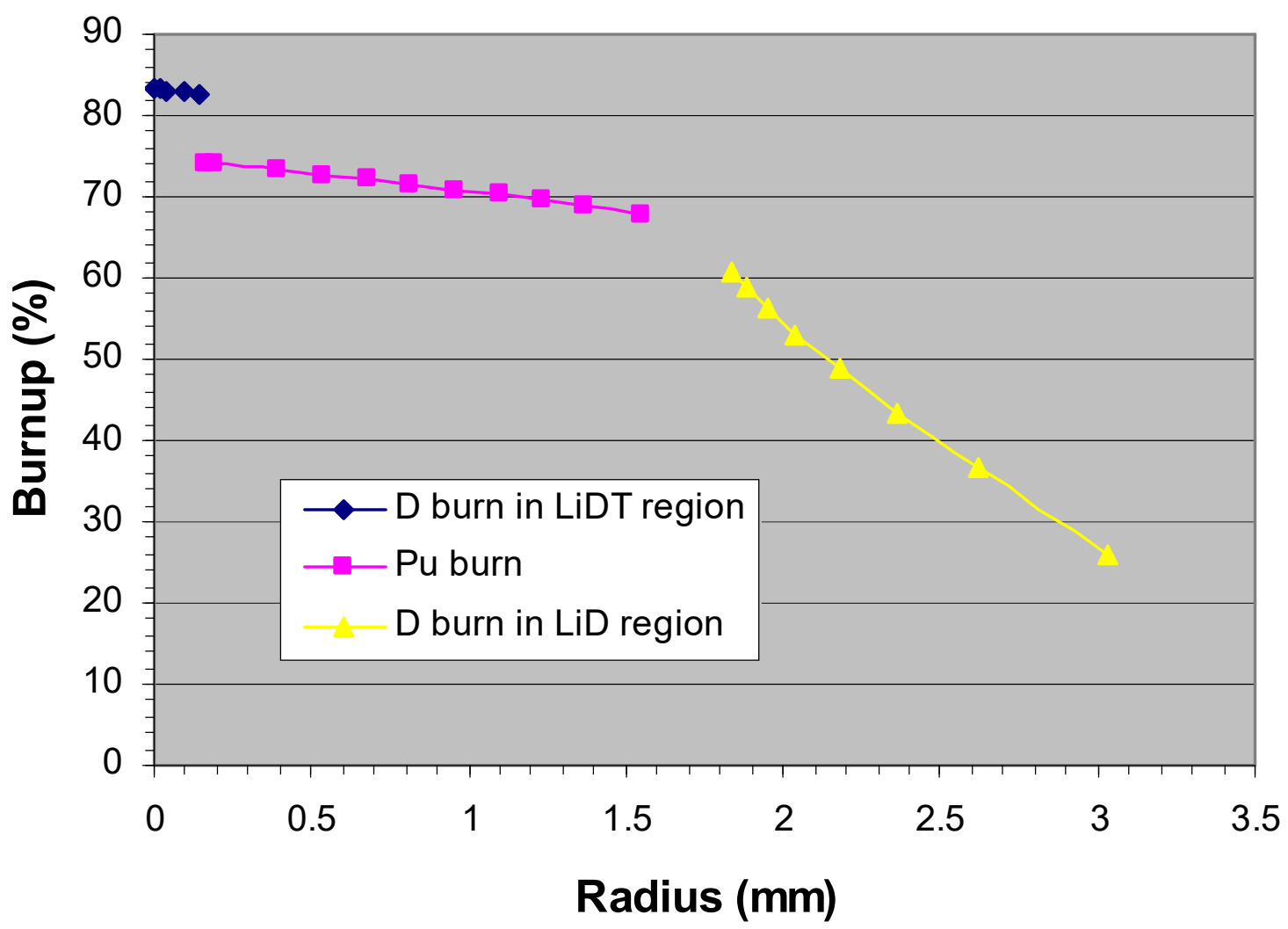

**Figure 20**: Final burnup of deuterium and plutonium.

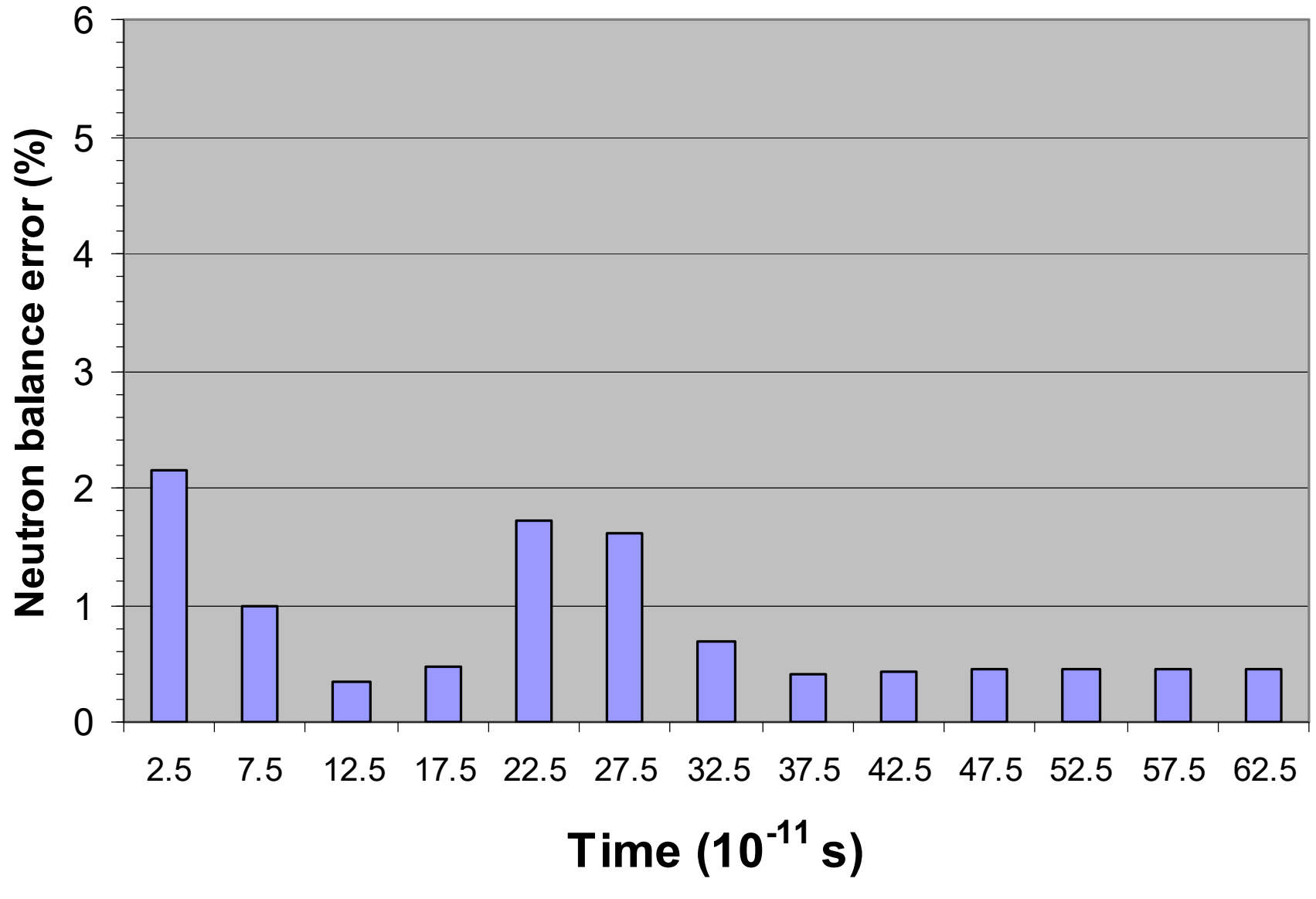

**Figure 21**: Error in neutron balance.

Below, we present the final inventory of the isotopes in all region and the final neutron balance (at time $t=68.3 \times 10^{-11}$s). The mixtures 12, 13, 15 and 16 correspond, respectively, to LiDT region, to Pu region, to main LiD fuel region and to U-nat region.

```
MIXTURE NUMBER: 13
REMAINDER MASS FRACTION OF Pu ISOTOPES RELATIVE TO THE TOTAL MASS OF THE MIXTURE AFTER DEPLETION
PU-239:0.258510D+00 PU-240:0.441509D-01 PU-241:0.320126D-02 PU-242:0.429029D-03

MIXTURE NUMBER: 16
REMAINDER MASS FRACTION OF U-235/U-236 ISOTOPES RELATIVE TO THE TOTAL MASS OF THE MIXTURE AFTER DEPLETION
U-235:0.602263D-02 U-236:0.000000D+00
FRACTION OF U-238 BURNED BY FISSIONS:0.498266D-01

MIXTURE NUMBER: 12
REMAINDER ATOMIC FRACTION OF DEUTERIUM IN THE MIXTURE:0.402562D-01
REMAINDER ATOMIC FRACTION OF TRITIUM IN THE MIXTURE:0.880792D-01
FRACTION OF D BURNED BY FUSIONS:0.838975D+00
REMAINDER ATOMIC FRACTION OF LI-6 IN THE MIXTURE:0.237839D-01
REMAINDER ATOMIC FRACTION OF LI-7 IN THE MIXTURE:0.439523D+00
FRACTION OF Li-6 BURNED BY n(He,T)Li-6 REACTION: 0.358925D+00
FRACTION OF Li-7 BURNED BY n(n,He,T)Li-7 REACTION: 0.505010D-01
TOTAL NUMBER OF FUSIONS: 0.146054D+19
FRACTION OF D-D FUSION REACTION TO THE TOTAL NUMBER OF FUSIONS: 0.210018D-01
FRACTION OF D-T FUSION REACTION TO THE TOTAL NUMBER OF FUSIONS: 0.977323D+00
FRACTION OF D-HE3 FUSION REACTION TO THE TOTAL NUMBER OF FUSIONS: 0.167473D-02

MIXTURE NUMBER: 15
REMAINDER ATOMIC FRACTION OF DEUTERIUM IN THE MIXTURE:0.277787D+00
REMAINDER ATOMIC FRACTION OF TRITIUM IN THE MIXTURE:0.200595D-01
FRACTION OF D BURNED BY FUSIONS:0.444426D+00
REMAINDER ATOMIC FRACTION OF LI-6 IN THE MIXTURE:0.374592D+00
REMAINDER ATOMIC FRACTION OF LI-7 IN THE MIXTURE:0.478220D-01
FRACTION OF Li-6 BURNED BY n(He,T)Li-6 REACTION: 0.167572D+00
FRACTION OF Li-7 BURNED BY n(n,He,T)Li-7 REACTION: 0.435599D-01
TOTAL NUMBER OF FUSIONS: 0.305407D+23
FRACTION OF D-D FUSION REACTION TO THE TOTAL NUMBER OF FUSIONS: 0.374253D+00
FRACTION OF D-T FUSION REACTION TO THE TOTAL NUMBER OF FUSIONS: 0.542891D+00
FRACTION OF D-HE3 FUSION REACTION TO THE TOTAL NUMBER OF FUSIONS: 0.828564D-01

 BALANCE OF NEUTRONS
    TOTAL NUMBER OF NEUTRONS CURRENTLY PRESENT IN THE SYSTEM=  0.2806D+22
    TOTAL NUMBER OF NEUTRONS LEAKED OUT FROM THE SYSTEM=  0.9855D+22
    TOTAL NUMBER OF NEUTRONS ABSORBED (CAPTURE+FISSION)=  0.1596D+23
    TOTAL NUMBER OF NEUTRONS PRODUCED BY ALL FISSIONS=  0.3536D+22
    TOTAL NUMBER OF NEUTRONS PRODUCED BY ALL FUSIONS=  0.2228D+23
    TOTAL NUMBER OF NEUTRONS PRODUCED BY ALL (N,2N) REACTIONS=  0.2673D+22
    PERCENTAGE ERROR=100xABS{1-[N(INITIAL)+N(PRODUCED)]/[N(SYSTEM)+N(ABSORBED)+N(LEAKAGE)]}=  0.4550D+00(%)
```

| N of $S_N$ | G | Energy ($10^3$ MJ) |
|---|---|---|
| 8 | 10 | 105.7 |
| 16 | 10 | 105.7 |
| 32 | 10 | 105.7 |
| 16 | $10^{(*)}$ | 96.0 |
| 16 | $7^{(*)}$ | 96.8 |

(*) without "upscattering" as explained in item 6.

**Table 3**: Energy produced by the microexplosion for different angular discretizations of neutron flux and, in the last two cases, by not considering "upscattering" in neutronic calculations.

## 8. Conclusion

We presented a numeric solution of the time-dependent neutron transport equation using a characteristic method, with the main objective of overcoming the negative flux problem that arises in some numeric scheme when the system is very opaque. In time dependent numerical solution this certainly occurs when the $1/v_g\Delta t$ term (that appears in the most common time discretization of time-dependent neutron transport equation) is much greater than total cross sections. This is worse when there is a combination of lower neutron velocities with very small time-steps, the latter occasionally imposed from outside.

The solution was proved in two applications involving symbiotic fission-fusion microsystems, when it is mandatory a full solution of the time-dependent neutron transport equation coupled to radiation-hydrodynamic and fission-fusion burnup equations.

The code used in the simulations also allows a solution of the time-dependent neutron transport equation using spatial interpolation of the fluxes in the meshes (as in diamond scheme). In the first application case, this solution produced good and similar results, but in the second more complex case it failed completely, generating very negative fluxes and increasingly divergent neutron balances.

Finally, very interesting results were obtained concerning the hybrid fission-fusion systems. One observation is that, although the Hansen-Roach cross sections yield good results in fast and intermediate fission systems, a better set of cross sections with a more refined and extended group structure in the higher energy range characteristic of fusion systems would be necessary (our artificial addition of an upper 10-15 MeV group only mitigate this fault). The incorporation of cross-sections of all fusion products is also recommended.